\documentclass[showpacs,amsmath,amssymb,amsfonts,pre,floatfix,twocolumn,superscriptaddress]{revtex4} 
\usepackage{graphicx}
\usepackage{bm}
\usepackage[german,english]{babel}

\begin{document}

\title{Correlation versus commensurability effects for finite bosonic systems in one-dimensional lattices}

\author{Ioannis Brouzos}
\email{ibrouzos@physnet.uni-hamburg.de}
\affiliation{Zentrum f\"ur Optische Quantentechnologien, Luruper Chaussee 149, 22761 Hamburg, Germany}

\author {Sascha Z\"ollner}
\email{zoellner@nbi.dk}
\affiliation{Niels Bohr International Academy, Niels Bohr Institute, Blegdamsvej 17, 2100 K\o benhavn, Denmark}

\author{Peter Schmelcher}
\email{Peter.Schmelcher@physnet.uni-hamburg.de}
\affiliation{Zentrum f\"ur Optische Quantentechnologien, Luruper Chaussee 149, 22761 Hamburg, Germany}

\date{\today}

\begin{abstract}
We investigate few-boson systems in finite one-dimensional multi-well traps covering the full interaction crossover from uncorrelated to fermionized particles. 
Our treatment of the ground state properties is based on the numerically exact Multi-Configurational Time-Dependent Hartree method.
For commensurate filling we trace the fingerprints of localisation, as the interaction strength increases, in several observables like reduced density matrices, fluctuations and momentum distribution. 
For filling factor larger than one we observe on-site repulsion effects in the densities and fragmentation of particles beyond the validity of the Bose-Hubbard model upon approaching the Tonks-Girardeau limit. 
The presence of an incommensurate fraction of particles induces incomplete localisation and spatial modulations of the density profiles, taking into account the finite size of the system.
\end{abstract}
 \maketitle

\section{Introduction}

Ultracold gases and Bose-Einstein condensates represent a highly controllable and rich many-body system \cite{pitaevskii03,pethick01}, 
appropriate for the investigation of several quantum phenomena. 
The possibility to accurately design the external forces employing static or laser fields, along with the tuning of the interaction strength by visiting Feshbach resonances,
allows us to prepare and explore weakly or strongly correlated quantum systems in almost arbitrary potential landscapes \cite{bloch08}. 

In particular, the flexible experimental toolbox of optical lattices enabled the investigation of the phase transition from a coherent superfluid (SF) to a Mott-insulator (MI) state with localized particles in each site \cite{greiner02},
 studied theoretically by the Bose-Hubbard model (BHM)  \cite{fisher89,jaksch98}. 
Further highlights include quantum phases like Bose glass and Mott-shells, occurring for disordered,
 confined and incommensurate systems \cite{batrouni02,rey05,kashurnikov02,buechler03,sengupta05,campbell06,luehmann09,damski03,rigol09,wessel04}. 

Tuning appropriately the laser beams it is possible to effectively decrease the dimensionality of the system and explore intrinsic phenomena of quasi one-dimensional (1D) systems,
 where the so-called confinement-induced resonances resulting from the change of the transverse length scale \cite{olshanii98} can be utilized to alter the interaction strength. 
This way, it became experimentally feasible to investigate weak to strongly interacting systems and especially the extreme case of the  Tonks-Girardeau (TG) gas,
 a state of strongly correlated bosonic matter intrinsically tied to 1D physics \cite{kinoshita04,paredes04}
 where bosons that repel each other with infinitely strong forces can be mapped to non-interacting fermions \cite{girardeau60}. 
Not only the special features of such a gas have attracted theoretical studies  \cite{girardeau01,vaidya79,rigol05}
 but also the crossover to this limit from 3D to 1D \cite{astrakharchik02,blume02b}.

While the analytical treatment of the interaction crossover in 1D from uncorrelated to TG-gas is possible for a homogeneous system with periodic \cite{lieb63a} or hard-wall \cite{hao06} boundary conditions,
 the exact study of bosonic systems exposed to specific trapping geometries is particular for a small number of particles. 
The harmonic trap \cite{busch98,arnold03,schmidt07,deuretzbacher06} and the double well \cite{busch03,murphy07,masiello05,zoellner06a,zoellner06b} are paradigm systems
 which unveil characteristic features of this crossover in confined geometries. 
Broadening and immergence of oscillations of the densities, as well as the evolution of the coherence properties have been examined in detail.

1D optical lattices are very special and appealing systems since they combine the physics in a lattice (phase transitions)
 with the special features of one dimension \cite{lin07,wei09,cazalilla03,cazalilla04,buechler03,rigol05,jiang07}. 
Localization, delocalisation as well as lattice imperfection effects have been studied for small ensembles \cite{luehmann08} showing interesting analogies with macroscopic phases. 
In particular momentum distributions, pair correlations and energy spectra give insight to the MI-SF transition with increasing lattice depth or unveil the effect of an incommensurate filling,
 while a Bose-glass phase emerges when breaking the lattice symmetry. 
The effect of a higher filling factor and the subsequent on-site fragmentation of particles in periodic 1D lattices has been explored in \cite{alon05b}. 
The authors investigate a commensurate case with two particles per site and distinguish between a phase of MI with unperturbed Wannier functions, 
and a second transition to a state with two fragmented orbitals on each site. 

In this article we cover the different interaction regimes from the  weakly correlated to the fermionization limit for a few-boson ensemble in a finite one-dimensional lattice.
We go far beyond the BHM regime where the change of the interactions is equivalent to the change of the lattice depth. 
Since commensurability is a key issue concerning the crossover, we examine representative few-body setups for commensurate and incommensurate filling factors. 
We demonstrate localization mechanisms by examining densities, particle number fluctuations and coherence loss in momentum distributions in the case of a unit filling factor. 
For higher commensurate filling we observe on-site repulsion effects such as broadening and oscillatory patterns in the densities which go beyond the Bose-Hubbard regime. 
Incommensurable setups on the other hand, show the possibility for spatial variations of observables, as a result of the correlations and the finite size of the system,
 in addition to partial delocalisation due to the incommensurate fraction of particles. 
The method which we use here is the Multi-Configurational Time-dependent Hartree (MCTDH)  \cite{meyer90,mctdhbook}
 which has proven to be efficient in treating bosonic systems both for static properties and dynamics (see Appendix) \cite{zoellner06a,zoellner06b,zoellner08b}.

The paper is organised as follows: In Sec. II we explain the setup, the analytically describable limits and the approximative methods. 
The results are presented in Sec. III for commensurate and Sec. IV for incommensurate filling. 
We summarise our findings and give an outlook in Sec. V. A short description of the numerical method (MCTDH) is included in the Appendix. 

\section{Setup and theoretical background}
\subsection{Model Hamiltonian}

The effective 1D Hamiltonian we consider reads:
\begin{equation}
 H=\sum_i^N h_i + \sum_{i<j} V_{\mathrm{int}}(x_i-x_j)
\end{equation} 
where the two-body interaction potential is delta-like $V_{\mathrm{int}}(x_i-x_j)=g_{1D}\delta(x_i-x_j)$. 
The coupling strength $g_{1D}$ depends on the scattering length $a_0$ and the oscillator length $a_{\perp}=\sqrt{\frac{\hbar}{M\omega_{\perp}}}$ which characterises the transverse confinement: 
$g_{1D}= \frac{2\hbar^2 a_0}{M a^2_{\perp}} (1-\frac{|\zeta(1/2)| a_0}{\sqrt{2} a_{\perp}})^{-1}$ \cite{olshanii98,zoellner06a}.
The one-body part of the Hamiltonian $h_i=\frac{p_i^2}{2M} + V(x_i)$ contains the 1D- lattice potential $V(x_i)=V_0 \sin^2(\kappa x_i)$,
 characterised by the depth $V_0$ and periodicity $d$ (distance between two successive minima) setting $\kappa=\pi/d$. 
In order to restrict the infinite potential $V(x_i)=V_0 \sin^2(\kappa x_i)$ to a finite number of sites $W$ and a length $L$, we impose hard-wall boundary conditions on appropriate position. 
In our calculations, for numerical convenience (keeping a standard grid), the length unit is $a_{\parallel}=L/10$ and the energy unit is chosen as $\hbar^2/Ma^2_{\parallel}$ setting also $\hbar=M=1$. 
The rescaled scattering strength is thus $g=\frac{a_{\parallel}M g_{1D}}{\hbar^2}$. 
In units of the recoil energy $E_R=\frac{\hbar^2 \kappa^2}{2M}$ it reads  $g'=\frac{g_{1D}}{dE_R}=\frac{2d}{a_{\parallel}\pi^2} g$. 

All the parameters are considered to be tunable almost at will in corresponding experiments. 
For the interaction strength $g$ we take representative values to cover the complete crossover to fermionization in all different cases of commensurability. 
We use a sufficiently large lattice depth ($V_0=7.0-20.0$ which is of the order of $4-20~E_R$ depending on $\kappa$) such that at least two single particle bands lie energetically below the continuum for reasons we explain later on. 
The way we render our system finite via imposing hard-wall boundaries does not restrict the generality of our treatment and results. 
In particular, independently of the specific experimental implementation (using multi-colour lattices and/or harmonic confinement), 
finite systems exhibit intrinsic features which differ from infinite lattices (periodic boundary conditions); 
the confined traps result always in spatial inhomogeneities of the densities which are absent in the periodic case. 
These inhomogeneities can be increased or manipulated if there is a harmonic confinement or a disordered surface. 
Although we are focusing here on the case of equal on site energies, we observe a rich behaviour that captures the main effects of a finite confined system. 
All our numerically exact calculations are performed by the MCTDH method (see appendix for a description). 
The following subsections referring to the extreme cases of no and infinite interactions, as well as the Bose-Hubbard model for weak interactions,
 are valid only in certain regimes which we address in the Secs. III, IV to explain the observations. 
Covering the complete interaction crossover from $g=0$ to $g \to \infty$ by MCTDH we are able to show the regimes of validity but also examine effects going beyond these models.

\subsection{Single Particle states}

In order to understand the limit cases of non- and infinitely- interacting particles (see next subsection), a discussion of the single particle states is necessary. 
Analytical expressions for the delocalised single-particle states, i.e. Bloch states, are available for periodic boundary conditions. 
For finite lattices, we use the tight-binding approximation assuming only a nearest-neighbour tunneling coupling term $J \propto -\int w_{s}(x) h_i w_{s+1}(x) dx$ between the sites $s$ and $s+1$,
 where $w_{s}(x)$ are the on-site localised Wannier states. 
Within this approximation, valid for a relatively deep potential, we express the Bloch states in terms of Wannier functions. 
The single-particle Hamiltonian written in a matrix form using the localised states basis is:

 \[\tilde{h}_i= \left(
\begin{array}{cccc}
\epsilon_{1} & -J & 0 & ..\\
-J & \epsilon_{2} & -J & ..\\
0 & -J & \epsilon_{3} & ..\\
.. & .. & .. & ..\end{array} \right) \]
where $\epsilon_{s}$ ($s=1,2,..,W$) are the on-site energies, which in our case are equal ($\epsilon_{1}=...=\epsilon_{W} \equiv \epsilon$).

\begin{figure}
\includegraphics[width=8.6 cm,height=4.0cm]{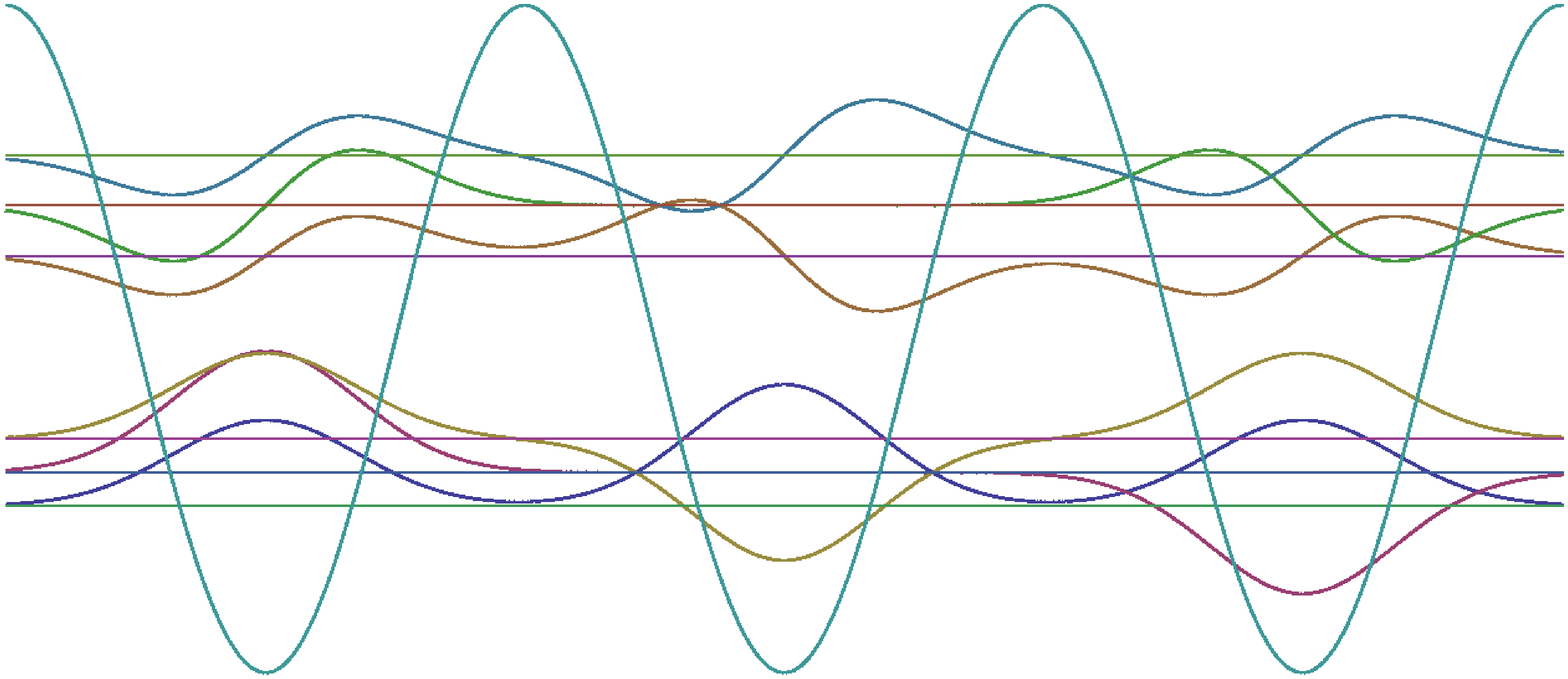}
\caption{(Colour online) Sketch of the finite three-well lattice and the corresponding single particle states for the first two bands.} \label{fig1}
\end{figure}

The hard-wall boundary conditions imply that there is no tunnel coupling between the first and the last lattice site as opposed to the periodic ones,
 where there is a single coupling for all sites. 
The resulting eigenvalues are:
$E_{q-1}=\epsilon-2J \cos(\frac{q\pi}{W+1})$ ($q=1,...,W$),
 and the eigenfunctions read:
\begin{equation}
|\varphi_{q-1}\rangle=\sqrt{\frac{2}{W+1}}\sum_{s=1}^W \sin \left(\frac{s q \pi}{W+1} \right) | w_s \rangle
\end{equation}
These lowest-band single-particle eigenstates are sketched in Fig.~\ref{fig1} for the triple well
 using the Gaussian approximation for the Wannier functions around the center of each well $\tilde{x}_s$ $\langle x|w_s \rangle = (\pi d^2)^{-1/4} e^{-(x-\tilde{x}_s)^2/2d^2}$:
$|\varphi_0 \rangle= \frac{1}{2}(|w_1 \rangle + \sqrt{2} |w_2 \rangle + |w_3 \rangle)$, 
$|\varphi_1 \rangle= \frac{1} {\sqrt{2}} (|w_1 \rangle -|w_3 \rangle)$, 
$|\varphi_2 \rangle= \frac{1}{2}( |w_1 \rangle -\sqrt{2} |w_2 \rangle + |w_3 \rangle)$, 
with corresponding energies $E_0=\epsilon-\sqrt{2}J, E_1=\epsilon, E_2=\epsilon+\sqrt{2}J$. 
Note that for the ground state the middle well is occupied with a larger amplitude compared to the two outer ones. 
For the states of the excited bands the harmonic oscillator orbitals of higher order can serve as localised functions to a rather good approximation. 
The computation of the eigenstates, done by numerical diagonalisation of the model Hamiltonian $h_i$ is in good agreement with this simple model $\tilde{h}_i$. 
In a recent paper \cite{rey05},
 the single particle states of the lowest band for the case of a lattice with a superimposed parabolic trap were derived analytically within the tight-binding approximation.

\subsection{Bose-Fermi map}

The non-interacting and the infinitely strongly interacting case of impenetrable bosons can be both addressed analytically,
 given that we can solve the corresponding single-particle problem. 
In the former case, we just have one orbital for the ground state (the lowest eigenstate of the single-particle Hamiltonian) where all bosons reside $\Psi_0=\phi_0^{\otimes N}$. 
In the Tonks-Girardeau limit, the Bose-Fermi map \cite{girardeau60} establishes the following isomorphy between the bosonic $\Psi_{\mathrm{TG}}$ and the non-interacting fermionic wave function
 $\Psi_{\mathrm{fer}}$: $\Psi_{\mathrm{TG}}(Q)= A(Q) \Psi_{\mathrm{fer}}(Q)$ where $A(Q)=\Pi_{i<j} \mathrm{sgn} (x_i-x_j)$ and $Q \equiv (x_1,...,x_N)^T$. 
In particular, the infinitely interacting bosonic ground state is simply the absolute value of the fermionic one.

The local densities are appropriate tools for analysing a many body state $|\Psi \rangle$:
 the one body density $\rho(x) \equiv \langle x | \hat{\rho}_1 | x \rangle$,
 diagonal kernel of the one-body density operator $\hat{\rho_1} \equiv \mathrm{tr}_{2,...,N} |\Psi \rangle  \langle \Psi|$,
 and the two body density $\rho_2(x_1,x_2) \equiv \langle x_1 x_2 | \hat{\rho_2} | x_1 x_2 \rangle$
 diagonal kernel of the two-body density operator $\hat{\rho_2} \equiv \mathrm{tr}_{3,...,N} | \Psi \rangle \langle \Psi |$. 
In the TG limit they are equal to the fermionic ones:
\begin{equation}
\nonumber
\rho^{TG}(x)=\sum_{\alpha=0}^{N-1} |\varphi_a(x)|^2
\end{equation}
\begin{eqnarray}
&\rho_2^{TG}(x_1,x_2)=&\frac{1}{N(N-1)}
\nonumber\\
&\sum_{0\leq \alpha<\alpha' \leq N-1} &|\varphi_{\alpha}(x_1) \varphi_{\alpha'}(x_2)-\varphi_{\alpha}(x_2)\varphi_{\alpha'}(x_1)|^2.
\nonumber
\end{eqnarray}
Accordingly the ground state energy in the fermionization limit is $E_{\mathrm{TG}}(N)=\sum_{\alpha=0}^{N-1} E_{\alpha}$.

\subsection{Bose-Hubbard model and beyond}

Bosons in optical lattices have bee mostly studied in the literature within the Bose-Hubbard model \cite{bloch08},
 employing the tight-binding and lowest band approximation. 
Apart from the coupling term $J$ and the on-site energies $\epsilon_s$ introduced in $\tilde{h}_i$,
 there is additionally on-site interaction with strength $U$ which reads for a delta contact potential  $U=g \int dx |w_{s}(x)|^4$:
\begin{equation}
\nonumber
\hat{H}_{BH}=-J \sum_{<s,s'>} \hat{a}^{\dag}_s \hat{a}_{s'} + \frac{U}{2} \sum_{s} \hat{a}^{\dag}_s \hat{a}^{\dag}_s \hat{a}_{s} \hat{a}_{s} + \sum_{s} \epsilon_s \hat{a}^{\dag}_s \hat{a}_s
\end{equation}
where $<s,s'>$ indicates the sum over nearest neighbours. Both parameters $J$ and $U$ can be tuned by the lattice constants $V_0$ and $d$. 
Increasing for example the laser intensity, the lattice becomes deeper, and thus the ratio $U/J$ increases substantially leading to the MI phase in the commensurate case:
 the particles save interaction energy by being localised in different wells and the tunneling to neighbouring sites is strongly suppressed. 
The SF phase on the other hand, is characterised by phase coherence and delocalisation of the particles. 
Quantum phase transitions at zero temperature are triggered by quantum fluctuations, and in this sense, we can examine in our few-body ensembles signatures of the MI and SF phase. 
We use as a measure of localisation, the local (on-site) particle number fluctuations:
\begin{equation}
{\Delta N_s}^2= \langle{n_{s}}^2\rangle- {\langle n_s \rangle}^2= N[\rho_{2_s}(N-1) + \rho_{1_s}(1-N\rho_{1_s})]
\end{equation} 
where $\rho_{1_s}=\int_s dx \rho(x)$ and $\rho_{2_s}=\int_s\int_s dx_1 dx_2 \rho(x_1,x_2)$ are the one- and two- body densities respectively, integrated over the lattice site $s$. 
The SF-MI transition is accompanied by decreasing and finally vanishing fluctuations $\Delta N_s$ as $U/J$ increases
 (corresponding in some sense to vanishing local compressibility in the MI phase \cite{batrouni02}). 
Another sign of the transition is the loss of coherence due to the localisation of the particles in individual sites. 
This can be observed in the momentum distribution, where the visibility of the interference peaks in the coherent SF phase is reduced,
 ending up with a smoothened incoherent profile in the MI regime \cite{greiner02}. 

For filling factors higher than one, $\nu \equiv \frac{N}{W}>1$, it is necessary to go beyond the simple BHM,
 which assumes unperturbed Wannier orbitals and is restricted to the lowest band,
 one has to include higher band effects \cite{luehmann08} to examine the fermionization and generally strong correlation effects. 
The Bose-Fermi map already indicates that the lowest band levels are not sufficient to accommodate a number of fermionized bosons larger than the number of wells. 
Let us briefly mention effective models relevant for the discussion of filling factors higher than one. 
A modulation of the Wannier functions to take into account on-site interaction effects has been proposed \cite{hazzard09,li06}
 or splitting into two orbitals into the same well \cite{alon05b}. 
Another recent few-body study suggests to optimise the BHM parameters such that they agree with the exact results specifically in the strongly interacting regime \cite{schneider09}. 
Besides this, the concept of {\it extended fermionization} valid in the extreme limit of BHM $U/J \to \infty$
 under the assumption that the particles occupy different layers of MI and SF character \cite{pupillo06}
 gives a valuable picture for situations where an incommensurate fraction of particles sits on a commensurate localised background. 
We emphasise that the following results are obtained by the numerically exact MCTDH method and we only refer to other models for explanation and comparison.

\section{Commensurate filling}

We explore here firstly the commensurate filling case showing the fingerprints of localization for filling factors $\nu \equiv \frac{N}{W}=1$ and $\nu=2$,
 the latter being particularly interesting for the study of on-site interaction effects with two particles per site.

\subsection{Filling factor $\nu=1$}

\paragraph{The spatial distribution of the particles}

\begin{figure}
     \includegraphics[width=8.6 cm,height=4.3cm]{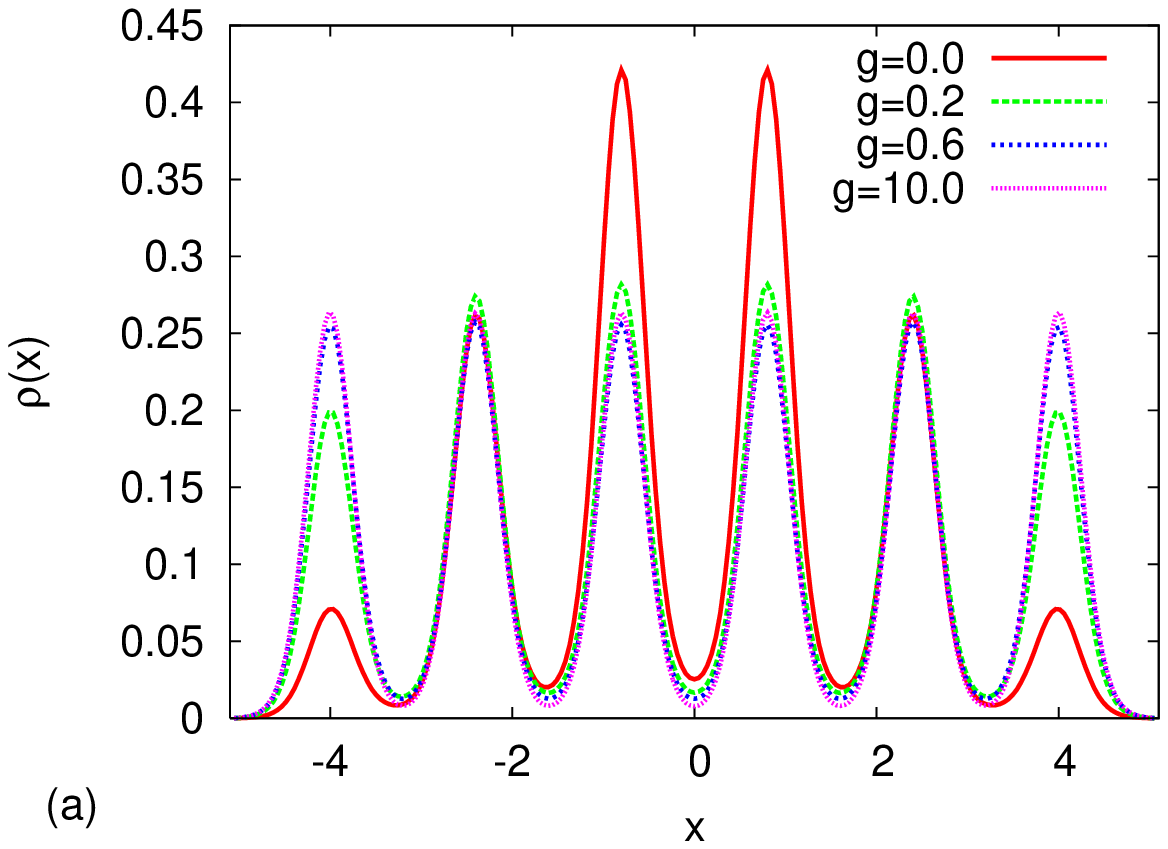} 
     \includegraphics[width=8.6 cm,height=4.3cm]{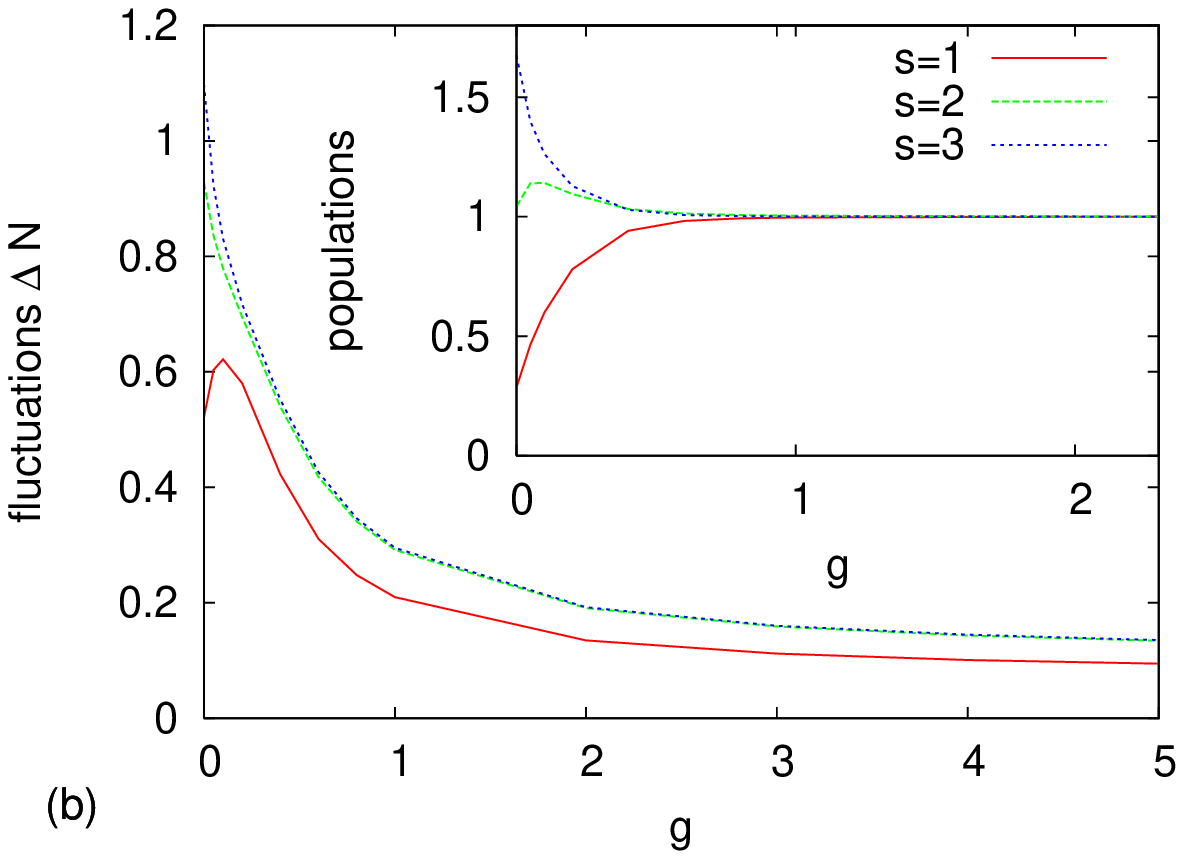} 

\caption{(Colour online) (a) One body-density $\rho(x)$ for 6 wells and 6 particles. 
Shown are 4 different values of $g$: non interacting ($g=0.0$), weakly interacting ($g=0.2,0.6$), Mott insulator-fermionization limit ($g=10.0$). 
(b) Particle number fluctuations as $g$ increases: red, green and blue lines correspond to the left 3 wells of the potential $s=1,2,3$ counting from the outermost one. 
(inset)  On-site population as $g$ increases.}
  \label{fig2}
 
\end{figure}

For the non-interacting ground state, the density of particles is larger for the middle sites and decreases as we go to the outer ones,
 illustrated here for 6 wells and 6 particles in Fig.~\ref{fig2}(a) for $g=0$. 
This occurs for setups with hard-wall boundary conditions but is generally characteristic of finite lattices (independent of the number of sites and particles),
 since the kinetic energy term renders the middle wells energetically more favourable. 
We remark that the particles reside all in the same single particle ground state which has indeed the maximal density in the middle:
 $\rho_1{_s} \propto |\sin \left(\frac{s \pi}{W+1} \right)|^2$ (see Eq. 2 and Fig.~\ref{fig1}). 

As the interaction strength increases (Fig.~\ref{fig2}(a)), we observe a gradual redistribution of the density which leads to equal population of all sites. 
Let us explain this observation in terms of the BHM which predicts a simple localisation process where each particle sits in one well to save interaction energy. 
In the Fock state representation $|N_1,N_2,...,N_W\rangle$, where each basis vector is parametrised by the occupation numbers for each site, with increasing $U/J$,
 the vector which has no double occupation $|1,1,1,... \rangle$ becomes the lowest eigenstate of the BHM Hamiltonian. 
As soon as the particles localise one per well, the increase of the interaction does not affect them anymore. 
In this fermionization limit their energy actually saturates to the fermionic value. 
Here $J$ is fixed (since there is a fixed depth $V_0=12.0$ corresponding to $6.2 E_R$ with $d=1.6$) and $U$ varies with $g$. 
In this special case $\nu=1$ and in general for $\nu \leq1$,
 the Bose-Fermi map does not hold only for the abstract state $\Psi_{\mathrm{TG}} \longleftrightarrow \Psi_{\mathrm{fer}}$ but also for the Wannier state \cite{cazalilla03,cazalilla04,buechler03}. 
In other words, the $U/J \to \infty$ limit of the BHM where each particles sits in one well $|1,1,1,... \rangle$, is here equivalent to the fermionization limit,
 where each particle occupies a single-particle level. Thus the BHM is valid in the case of $\nu=1$ for the whole range of interactions. 
In Fig.~\ref{fig2}(b) (inset) we plot the population of each well, $N \rho_{1_s}$, as a function of the interaction strength. 
While for the outer ($s=1$) and middle ($s=3$) wells the population evolves monotonically to the final value,
 the intermediate wells (see $s=2$) act as a 'carrier' of particles and thus their population  may exceed for some intermediate interaction strength ($g\approx 0.1$) their final one.

\paragraph{Diminishing fluctuations as a sign of localization}

The uniformisation of the population with increasing interaction strength is a pre-signature for the localization. 
A more accurate measure indicative for this mechanism, is the particle number fluctuations (Eq. 3) which decrease substantially as the interaction increases (Fig.~\ref{fig2}(b)) . 
Of course, a lower $J$, corresponding to a deeper lattice, enhances the localisation process. 
For a very deep lattice $U/J \to \infty$ we expect them to vanish completely, while in our case they saturate to a small value. 
The rather shallow depth of our lattice permits an occupation of the interwell space with a non-zero overlap between particles sitting in neighbouring sites. 
Including this area into the integrations of the one- and two- body density (Eq. 3) we end up with a small contribution to the fluctuations even in the strongly interacting limit. 
Note also that the middle wells ($s=3$), which can 'lose' population with respect to both sides, keep on having larger fluctuations than the outer ones ($s=1$).  

In Fig.~\ref{fig2}(b) we observe that the fluctuations converge to a constant value for $g\approx 3.0$. 
This convergence happens for values of $g$ significantly larger than those where the uniform distribution of the population is achieved ($g=0.6$ Fig.~\ref{fig2}(b) inset). 
Technically speaking, particle-hole excitations like  $|2,0,1,... \rangle$, $|0,2,1,... \rangle$ contribute for $g\approx 0.6$ resulting in an equal site distribution
 but without forming a localised state $|1,1,..,1 \rangle$. 
Perfect localisation occurs only when the latter vector is the eigenstate of the system suppressing all other contributions. 

Let us point out here the effect of the hard wall in comparison with periodic boundary conditions. 
In our confined system for weak interactions,
 the Fock states with less occupation in the outer wells outweigh those with less occupation in the center and therefore we observe an imbalance of the sites population. 
In the case of periodic boundary conditions even the non-interacting state possesses a uniform population of the sites due to symmetry. 
Hence, what really happens in the latter case within the transition from weak to strong interactions,
 is only a reduction of fluctuations by going to the state $|1,1,1,... \rangle$, and not a redistribution of the site populations as in our case.

\paragraph{Two-body correlations}

\begin{figure}
  \begin{center}
    \begin{tabular}{cc}
       \includegraphics[width=4.0 cm,height=4.0 cm]{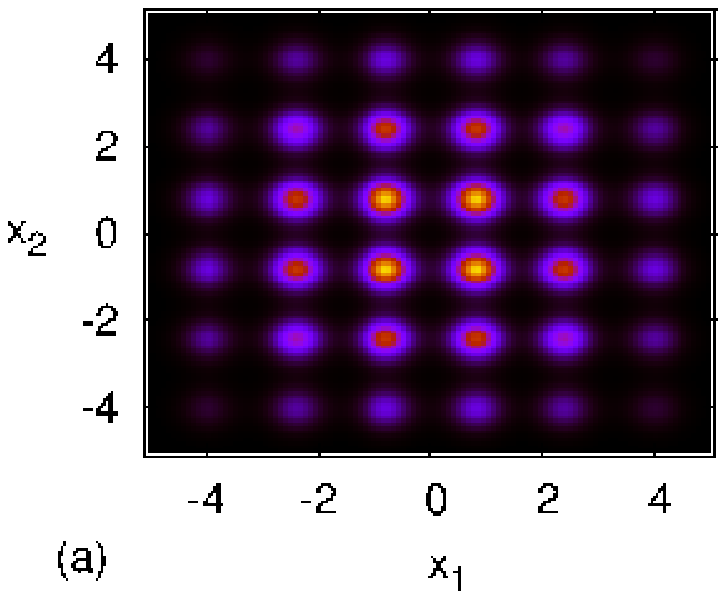} &
       \includegraphics[width=4.0 cm,height=4.0 cm]{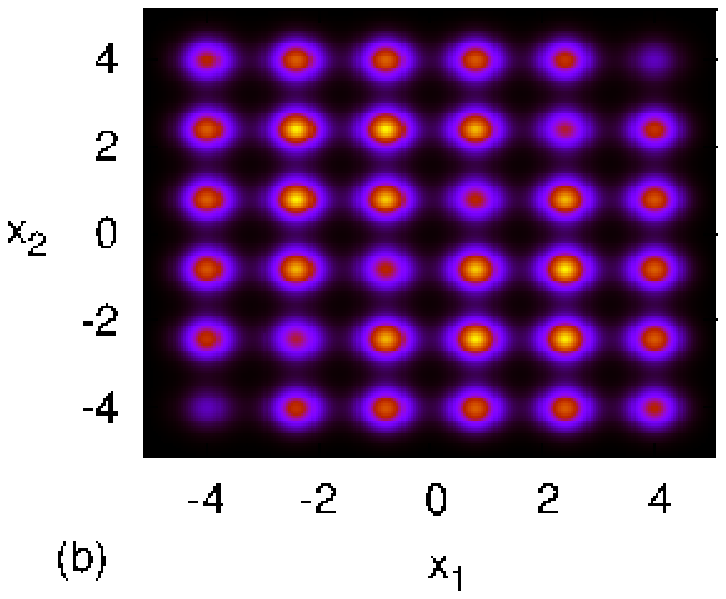}\\
       \includegraphics[width=4.0 cm,height=4.0 cm]{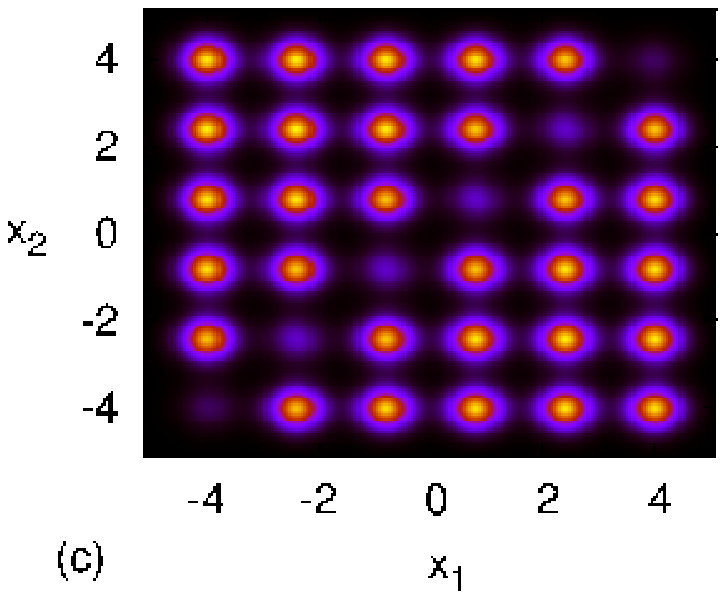} &
       \includegraphics[width=4.0 cm,height=4.0 cm]{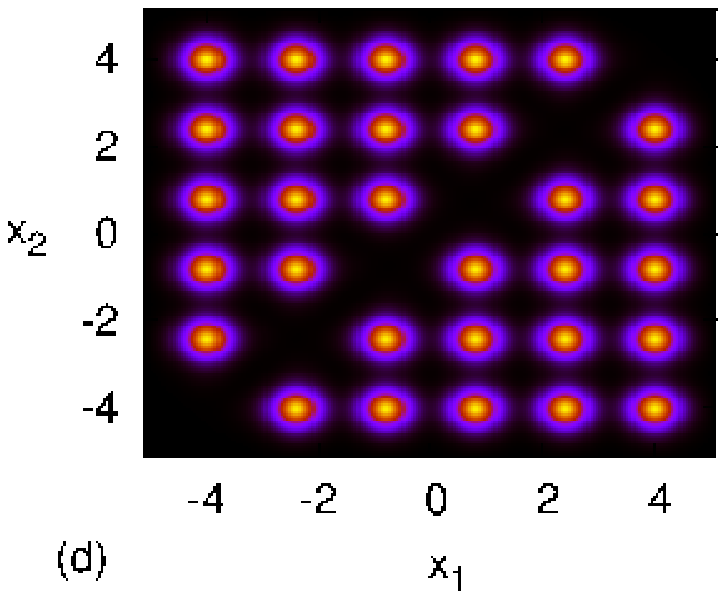} \\
    \end{tabular}
\caption{(Colour online) Two body-density $\rho_2(x_1,x_2)$ for 6 wells and 6 particles. 
Shown are 4 values of the interaction strength (a) $g=0.0$, (b) $g=0.2$, (c) $g=0.6$, (d) $g=10.0$.}
  \label{fig3}
  \end{center}
\end{figure}

In the two-body density $\rho_2(x_1,x_2)$ (Fig.~\ref{fig3}), starting from $g=0$ and for increasing interactions ($g=0.2$),
 the diagonal peaks indicating double occupation (especially the middle) fade out, while the off-diagonal ones are amplified. 
Note that in the regime where equal distribution over the sites has been achieved ($g=0.6$),
 the diagonal of the two body density has not yet been fully emptied,
 a sign that some double occupation is still present and thus a true localised state with unit filling is not yet reached. 
As $g\to \infty$ the repulsive forces totally prohibit double occupation and lead to a complete depletion of the diagonal ($g=10$). 
As a result, if one particle sits in one well, any second one distributes itself equally over the other wells, but has zero probability to be in the same well.

\paragraph{Non-local correlations}

We will now cast light on the system from the perspective of the non-local properties specifically the off-diagonal kernel of the one-body density matrix
 $\rho_1(x,x')\equiv \langle x|\tilde{\rho}_1|x' \rangle$ and the momentum distribution. 
While the local properties in the TG-limit have exactly the fermionic profile, the bosonic permutation symmetry plays a significant role for the coherence properties. 
The off-diagonal behaviour of  $\rho_1(x,x')$ as $|x-x'| \to \infty$ is a measure of coherence \cite{Yang62}:
 it indicates non-vanishing off-diagonal long range order (ODLRO) in infinite homogeneous systems. 
In our finite setups though there can be no true ODLRO, so the term \emph{coherence} refers here to the off-diagonal parts of $\rho_1(x,x')$ showing short and long range one-particle correlations.

\begin{figure}
  \begin{center}
    \begin{tabular}{cc}
      \includegraphics[width=4.0 cm,height=4.0 cm]{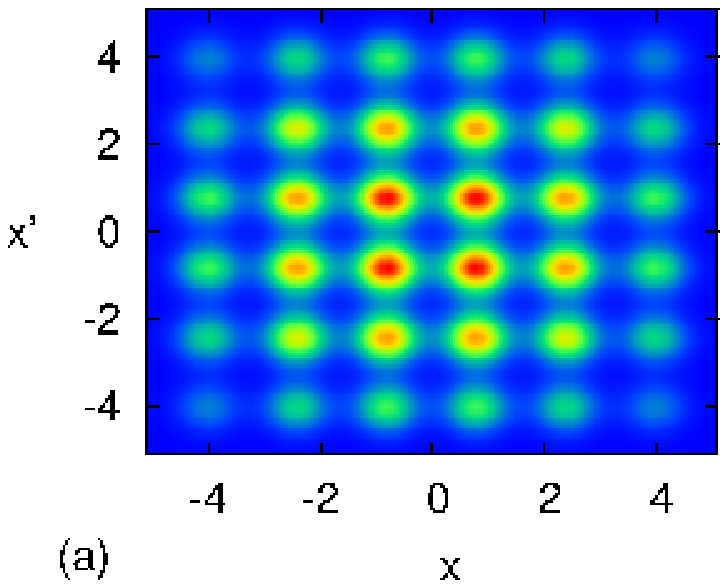} &
      \includegraphics[width=4.0 cm,height=4.0 cm]{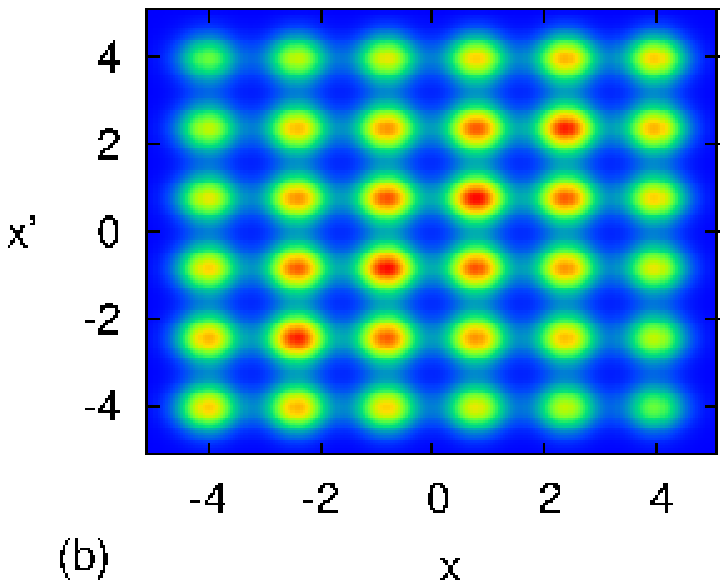} \\
      \includegraphics[width=4.0 cm,height=4.0 cm]{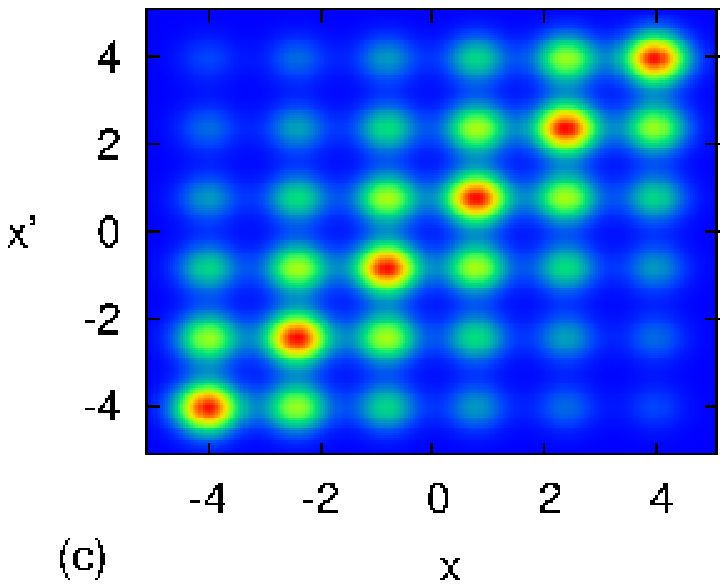} &
      \includegraphics[width=4.0 cm,height=4.0 cm]{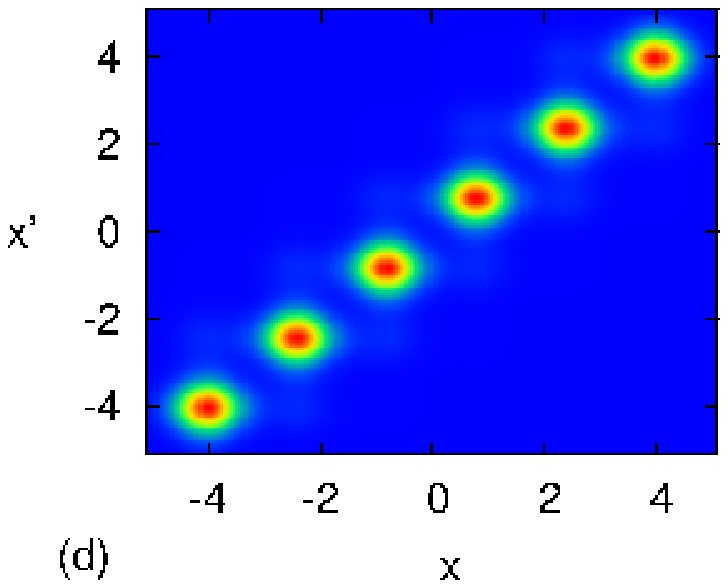} \\
    \end{tabular}
\caption{(Colour online) Off diagonal one-body density matrix $\rho_1(x,x')$ for 6 wells and 6 particles. Shown are 4 values of the interaction strength (a) $g=0.0$, (b) $g=0.2$, (c) $g=0.6$, (d) $g=10.0$.}
  \label{fig4}
  \end{center}
\end{figure}

For zero interactions (see Fig.~\ref{fig4} $g=0.0$) we observe that the off-diagonal spots fade out for an increasing distance from the center as expected for a confined system. 
For weak interactions ($g=0.2$), the off-diagonal contributions -especially the remote ones- become more pronounced, along with the outer wells on the diagonal. 
This is a common feature for finite setups, being accompanied by the initial density redistribution. 
The particles, redistribute all over the lattice such that they reduce the interaction energy, and as long as they stay in one orbital (see explanation later on),
 increase the correlations all over the space. 
As the interaction strength increases further ($g=0.6$), the localisation on discrete sites gradually destroys the ODLRO . 
However, some short range coherence persists as the distribution of the particles becomes uniform ($g=0.6$)
 and dies out for a stronger value of the interaction strength ($g=10.0$). 
This matches with our previous observation,
 that the diminishing of the fluctuations and coherence in the MI phase does not necessarily coincide with the appearance of a uniform population.

\paragraph{Momentum distribution}

The off-diagonal part of the one-body density matrix is not an observable but it is indirectly accessible via time-of-flight measurements \cite{bloch08}, which yield the momentum distribution $\tilde{\rho}(k)$:
\begin{equation}
\tilde{\rho}(k)=2\pi \langle k|\rho_1|k \rangle =\int dx \int dx' e^{-ik(x-x')} \rho_1(x,x').
\end{equation}

\begin{figure}

     \includegraphics[width=8.6 cm,height=4.3cm]{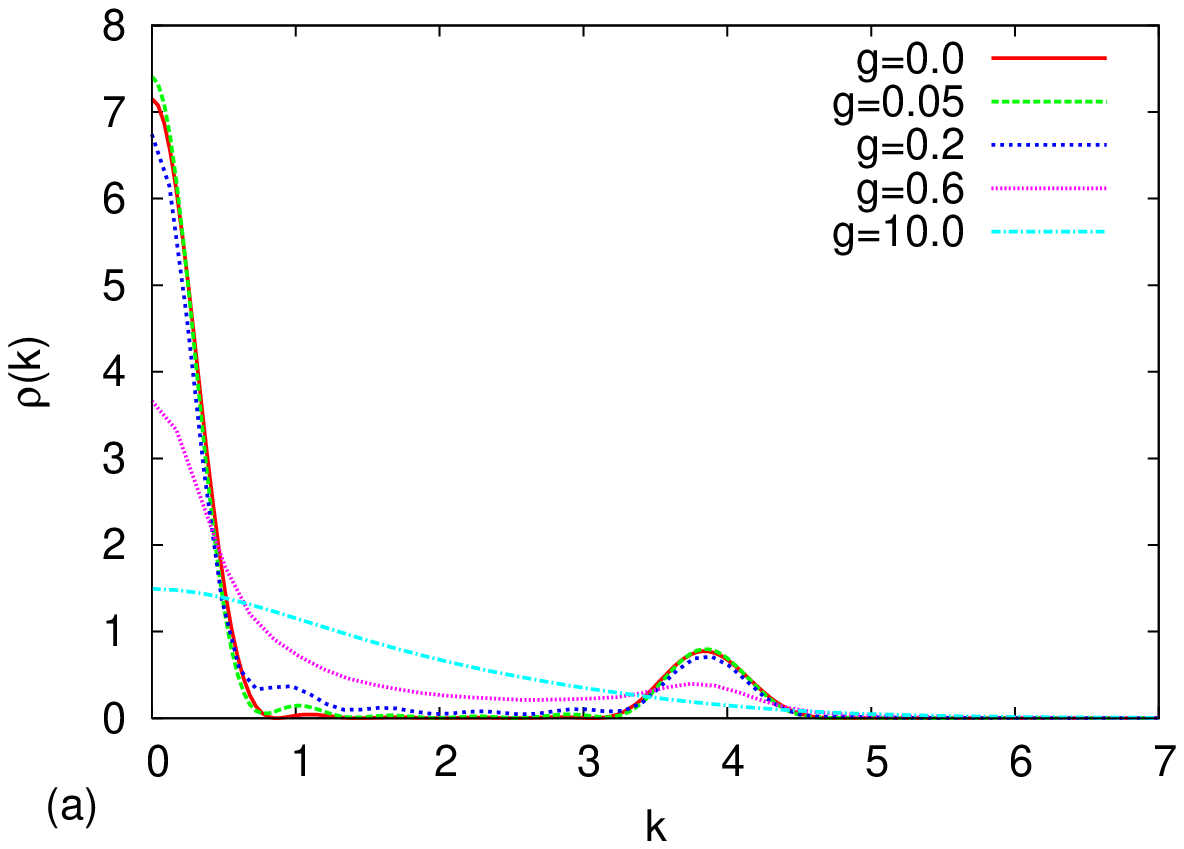} 
     \includegraphics[width=4.2 cm,height=4.2cm]{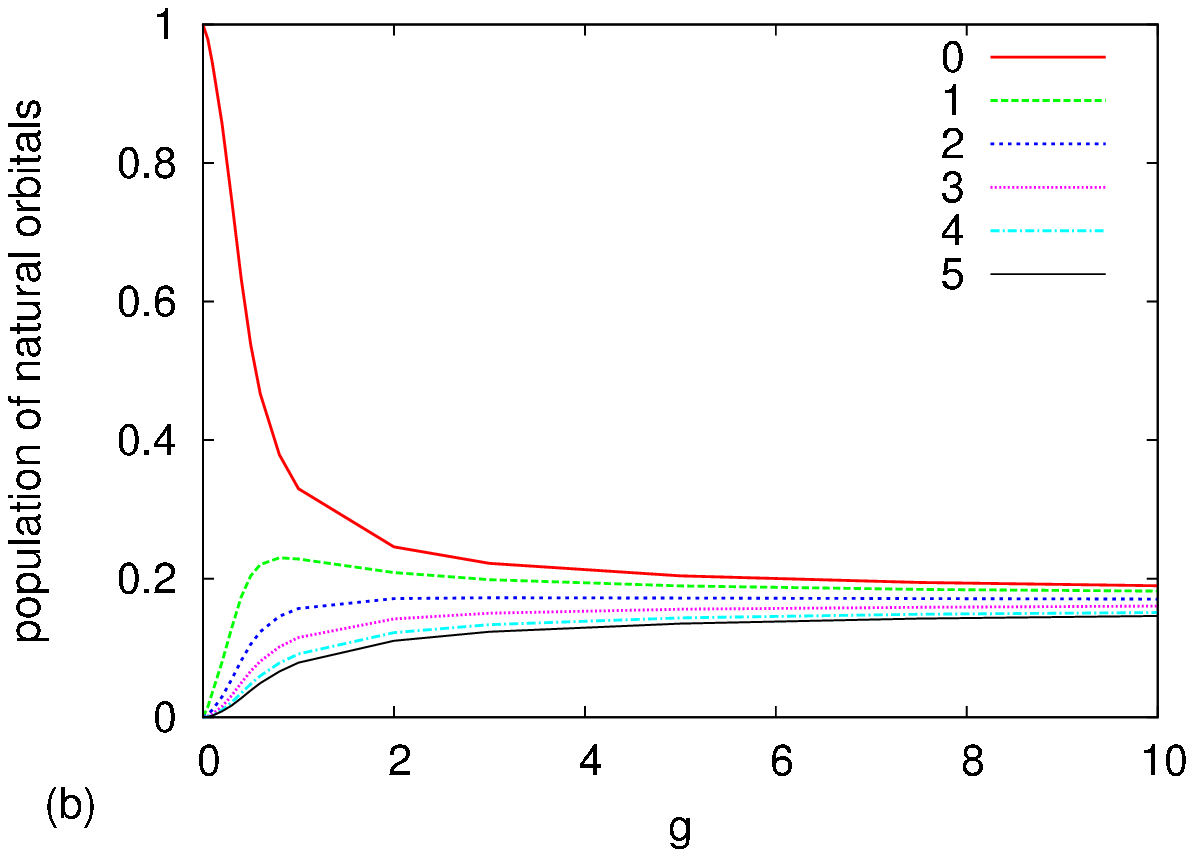} 
     \includegraphics[width=4.2 cm,height=4.2cm]{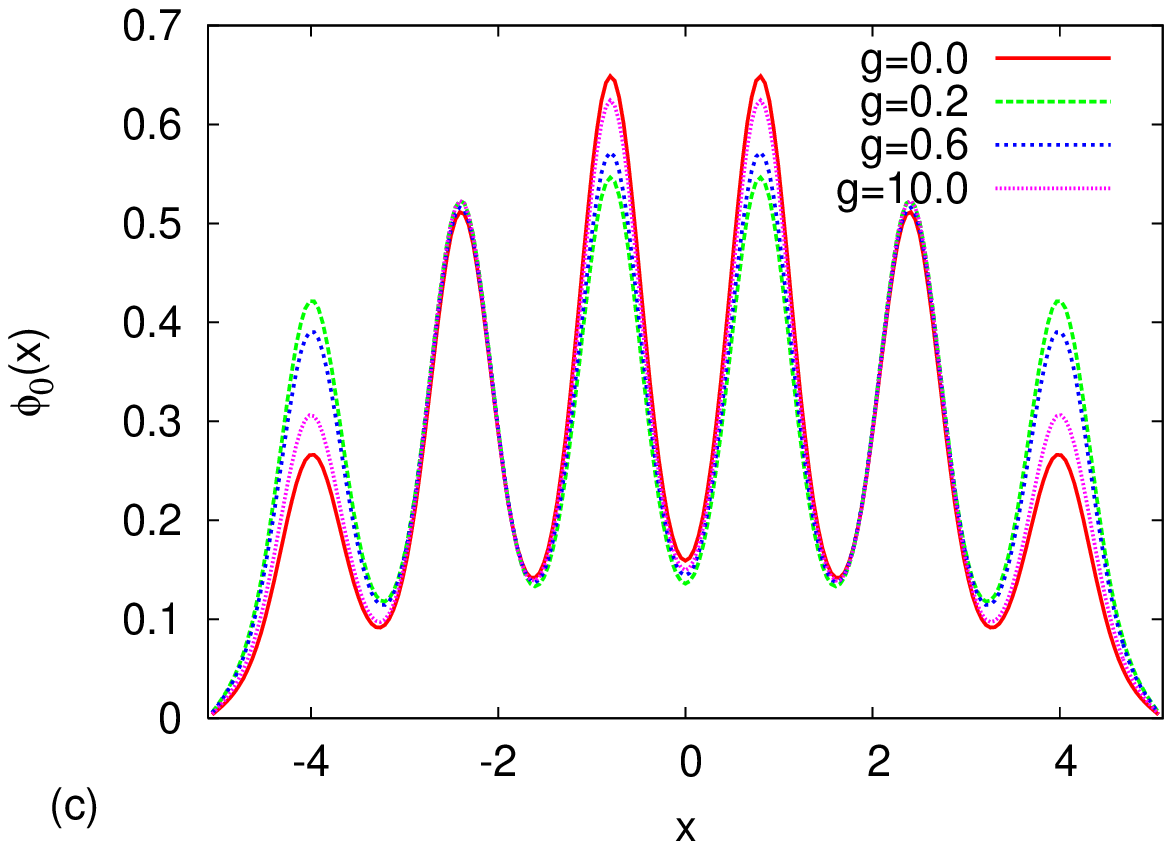} 

\caption{(Colour online) (a) Momentum distribution for 6 wells and 6 particles. Shown are 5 different values of $g$: non interacting ($g=0.0$), weakly interacting ($g=0.05,0.2,0.6$), Mott insulator-fermionization limit ($g=10.0$). (b) Population of the natural orbitals as a function of the interaction strength. (c) Profile of lowest natural orbital for several values of the interaction.}
  \label{fig5}
\end{figure}

In Fig.~\ref{fig5}(a) we observe that $\rho(k)$ exposes a rich pattern for $g=0$ with Bragg peaks near the reciprocal lattice vector ($k=a^{\ast}=\frac{2 \pi}{d}$ $d=1.6$)
 which is gradually smeared out as the interaction strength increases. 
The central peak corresponding to the ODLRO is high but gets even slightly higher for small interactions ($g=0.05$),
 matching with our observation in $\rho_1(x,x')$ that the remote off-diagonal humps increase. 
In this SF coherent regime ($g=0.0, 0.05, 0.2$) there occur also minor dips at the points $k_m=(m/W)a^{\ast}$,
 due to the suppression of standing waves with odd parity and wavelengths $\lambda_m=\frac{W d}{m}$ ($m=1,2,...$) resulting from the confinement,
 since the ground state possesses an even parity \cite{luehmann08}. 
For $g=0.6$ a value corresponding to uniform distribution in the density, the central peak is substantially lowered,
 which is a typical sign of coherence loss because of localisation,
 but the lattice geometry fingerprints (side peaks) in the momentum profile do not vanish completely implying again imperfect localisation. 
As the interaction increases further ($g=10.0$), $\rho(k)$ goes to a smoothened gaussian-like profile with complete destruction of the superfluid interference pattern
 and the visibility of the associated peaks \cite{greiner02,luehmann08}. We end up with a 'MI-type' incoherent state.

\paragraph{Fragmentation analysis via natural orbitals}

The complete information in the one-body level is given by the spectral decomposition of $\tilde{\rho}_1 \equiv \sum_{l=0} n_l |\phi_l \rangle  \langle \phi_l|$,
 where the relative populations $n_l$ serve as a measure of fragmentation into effective single particle states $\phi_l$ (natural orbitals). 
In \cite{penrose56}, a criterion for the a {\it non-fragmentated} condensed state was introduced exactly by demanding the highest such occupation $n_0$ to be very close to one.

For zero interactions, all the particles reside in the lowest (ground state) natural orbital (see Fig.~\ref{fig5}(b) $g=0.0$). 
Increasing the interaction they gradually {\it fragment} into the first $N=W=6$ orbitals. 
Thus, in this case ($\nu=1$) the effective description through the lowest-band single-particle states holds. 
The population of each orbital in the fermionization limit is not exactly $1/N$
 as we would naively expect from a mapping to non-interacting single-particle states according to Girardeau's theorem (for fermions $n_l=1/N$ for $l=0,N-1$). 
This is because the natural orbitals are effectively modulated single particle states originating from the spectral decomposition of $\tilde{\rho}_1$,
 and this modulation accounts for interaction effects. 
The lowest orbital $\phi_0$, for example (see Fig.~\ref{fig5}(c)), is broadened for weak interactions ($g=0.2$), following the evolution of the one-body density. 
The latter fact validates the use of the Gross-Pitaevkskii mean-field treatment for weak interactions
 where all the particles are assumed to reside in one variationally modulated orbital. 
The modulation of this dominant orbital is responsible for the initial extension of the off-diagonal range in $\rho_1(x,x')$
 and consequently leads to an increase of the central peak in $\rho(k)$ (see Fig.~\ref{fig5}(a)). 
The {\it fragmentation} observed for higher interactions (see Fig.~\ref{fig5}(b)), is beyond the regime of validity of the Gross-Pitaevskii equation,
 and due to the admixing of higher orbitals coherence is destroyed. 
This may be compared with the cases of the double well and the harmonic trap \cite{zoellner06a}:
 in the double well, the interaction immediately bridges the gap within the lowest-band doublet and thus destroys the coherence. 
Note that the ground state orbital in case of the double-well has already an equal distribution for the two wells
 and thus there is no dramatic flattening of the density due to the interactions as in the case of more wells examined here. 
In the harmonic trap there is an initial extension of the long-range order exactly for the same reason as here,
 i.e., the ground state orbital is broadened and holds the main population (the gap between the ground and the excited state is relatively large here). 
For very strong interactions (Fig.~\ref{fig5}(c) $g=10.0$),
 the profile of the orbitals tends to return to the non-interacting one indicating the validity of the Bose-Fermi map in the TG-limit.

\subsection{Filling factor $\nu=2$}

\begin{figure}

      \includegraphics[width=8.6 cm,height=4.3cm]{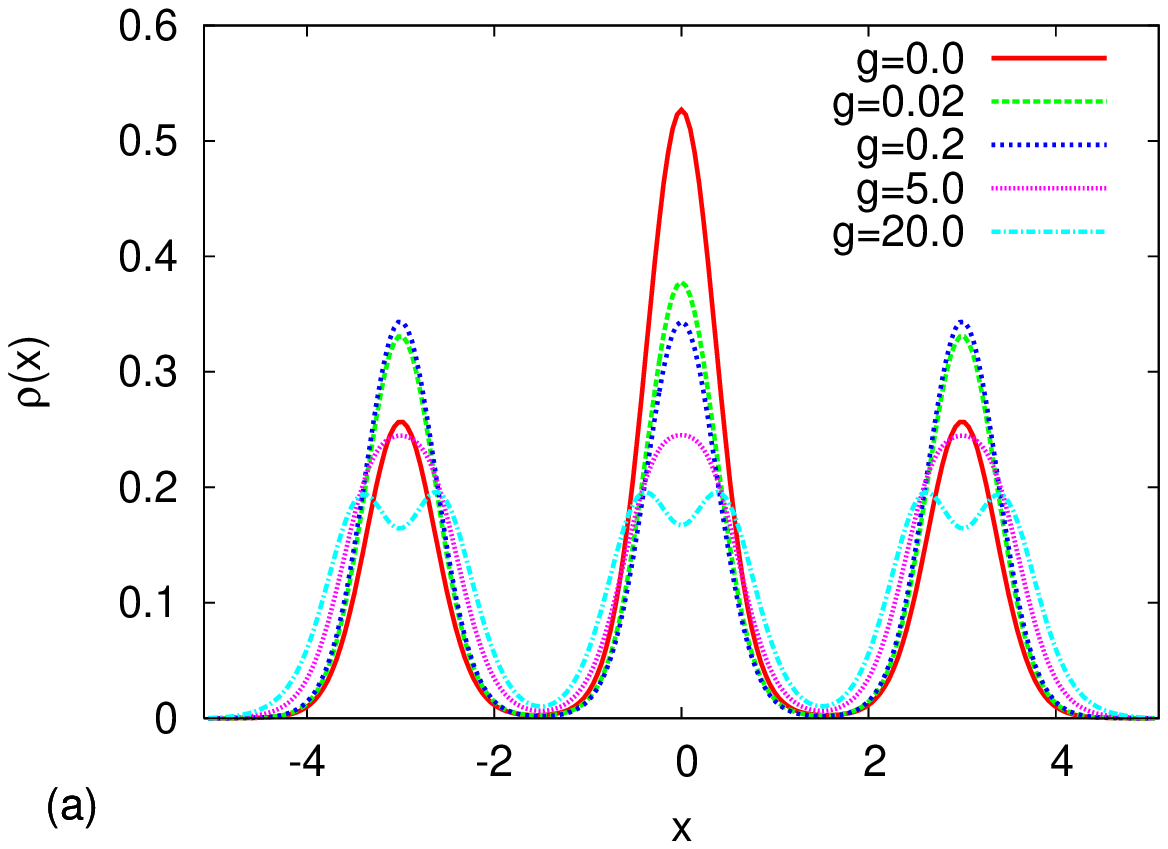} 
      \includegraphics[width=8.6 cm,height=4.3cm]{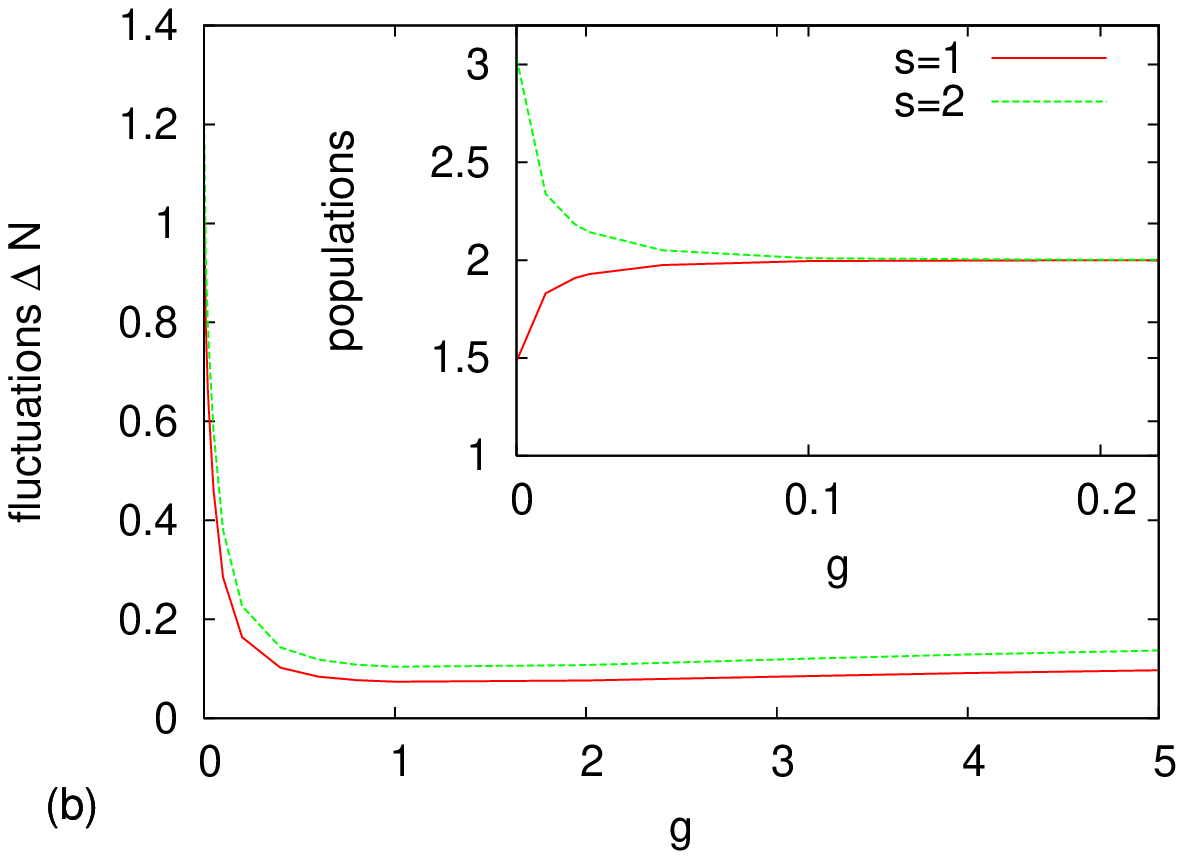} 

\caption{(Colour online) (a) $\rho(x)$ for 3 wells and 6 particles. Shown are 5 different values of $g$: non interacting ($g=0.0$), weakly interacting ($g=0.02$, $g=0.2$), and on-site fermionization crossover ($g=5.0$, $g=20.0$). (b) Particle number fluctuations as g increases  for the left ($s=1$) and the middle ($s=2$) well. (inset) On-site populations.}
  \label{fig6}

\end{figure}

In the case of $\nu=2$, here examined for 6 particles in 3 wells, we expect on-site interaction effects to be important because of the higher number density of particles. 
The one-body density redistributes resulting to an equal population for all wells with increasing repulsion (see Fig.~\ref{fig6}(a) $g= 0.2$). 
In terms of BHM, we have a formation of a 'Mott state' of 2 particles per site residing in unaltered Wannier orbitals
 as the vector $|2,2,2,... \rangle$ becomes eigenstate of the Hamiltonian for $U/J \to \infty$.

After this localisation into pairs is achieved, on site interaction effects become apparent in the one-body density beyond BHM. 
The well-studied and analytically solvable \cite{busch98} fermionization pathway of 2 particles in an individual harmonic trap arises on each site. 
In particular, there is a broadening of the one-body density in each well resulting from the increase of the onsite repulsion (Fig.~\ref{fig6}(a) $g= 5.0$). 
For even stronger interactions ($g= 20.0$), we observe the formation of two density maxima per site . 
These patterns arising in the few body setups for the strongly interacting regime have been examined for the harmonic trap \cite{schmidt07,deuretzbacher06,zoellner06a},
 the double well \cite{murphy07,yin08,mueller06,zoellner06a,zoellner06b} and 1D lattices with periodic boundary conditions \cite{alon05b}. 
In the latter case the authors distinguish two phases: first the localisation into pairs (BHM regime),
 and second fragmentation of each pair into two orbitals in the same well, which is consistent with our exact results. 
We underline that the on site interaction effects, go beyond the validity of the BHM, because higher band contributions need to be taken into account. 
In terms of the Bose-Fermi map, for $\nu=2$ all the levels of the first two bands are occupied in the fermionization limit. 
The upper band (see Fig.~\ref{fig1}) involves excited functions with one node per site. 
The combination of these functions with the lowest band ones, gives the observed wiggled fermionization profile with a local minimum in the middle of each well. 
Modulation of the Wannier functions has also been proposed to include on-site two-particle interaction effects \cite{hazzard09}. 
The particle number fluctuations (Fig.~\ref{fig6}(b)) together with the populations (Fig.~\ref{fig6}(b) inset) illuminate the whole process:
 first the particles distribute homogeneously ($g=0.0-0.1$),
 then localise and the fluctuations tend to a low value ($g \approx 0.1-1.0$) as expected for commensurate filling in the MI phase,
 and finally each pair of particles exhibits an individual on-site two-body crossover as described above.

\begin{figure}
  \begin{center}
    \begin{tabular}{cc}
      \includegraphics[width=4.0 cm,height=4.0 cm]{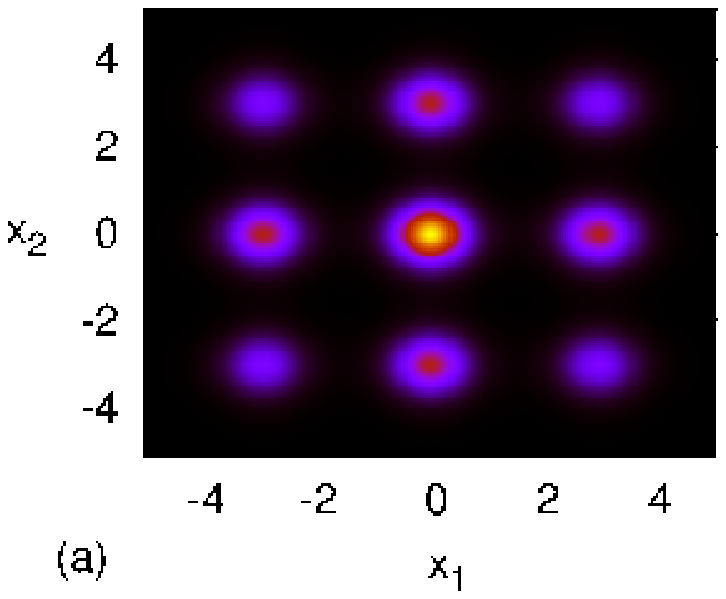} &
      \includegraphics[width=4.0 cm,height=4.0 cm]{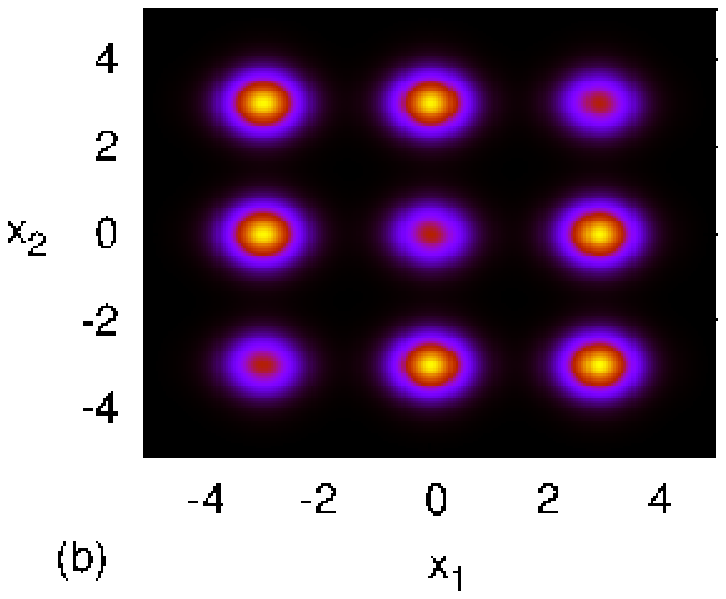} \\
      \includegraphics[width=4.0 cm,height=4.0 cm]{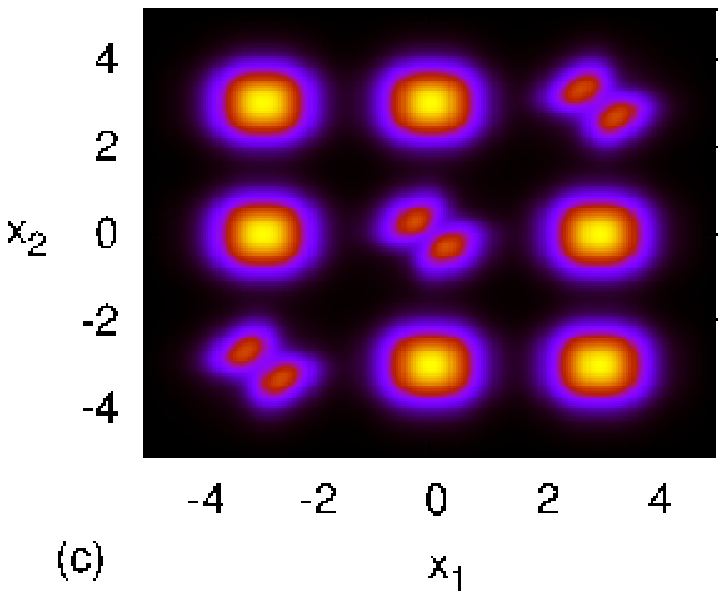} &
      \includegraphics[width=4.0 cm,height=4.0 cm]{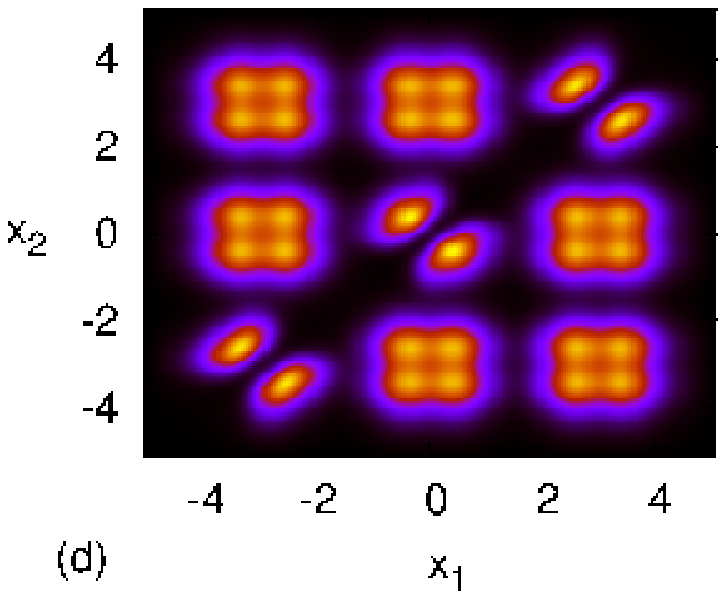} \\
    \end{tabular}
\caption{(Colour online) $\rho_2(x_1,x_2)$ for 3 wells and 6 particles for (a) $g=0.0$, (b) $g=0.2$, (c) $g=5.0$, (d) $g=20.0$.}
  \label{fig7}
  \end{center}
\end{figure}

In the two-body density $\rho_2(x_1,x_2)$ the diagonal contribution is reduced with increasing value of $g$ (see Fig.~\ref{fig7} $g=0.2$). 
In fact the maxima on the diagonal acquire half the population of any off-diagonal one, which matches with the pair localisation process leading to an equal population of all sites. 
For stronger interaction  (see Fig.~\ref{fig7} $g=5.0$) the formation of a {\it correlation hole} at $x_1=x_2$ in the diagonal occurs. 
This is an inherent two-body effect of the on-site fermionization process which is smoothened out in the integrated one-body density. 
The 'incomplete' correlation hole is a significant characteristic of $\nu>1$ filling factors in general,
 as the particles have no chance to be in completely different wells and thus are obliged to minimise their overlap on the same site (diagonal). 
The correlation hole is a very prominent two-body effect and occurs even for interactions interaction strengths (at $g \approx 0.6$)
 where the on-site broadening and maxima in the one-body density are not yet pronounced. 
Nevertheless, the formation of the correlation hole begins after the localisation of two particles per well is established. 
Also visible in Fig.~\ref{fig7}, is a broadening (for $g=5.0$) and a fragmented pattern (for $g=20.0$), which appear in the off-diagonal.

\begin{figure}

      \includegraphics[width=4.0 cm,height=4.0cm]{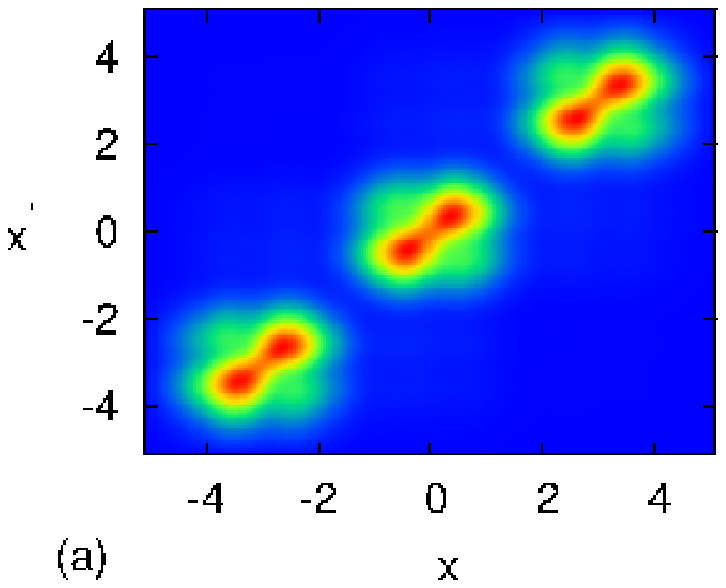} 
       \includegraphics[width=4.3 cm,height=4.0cm]{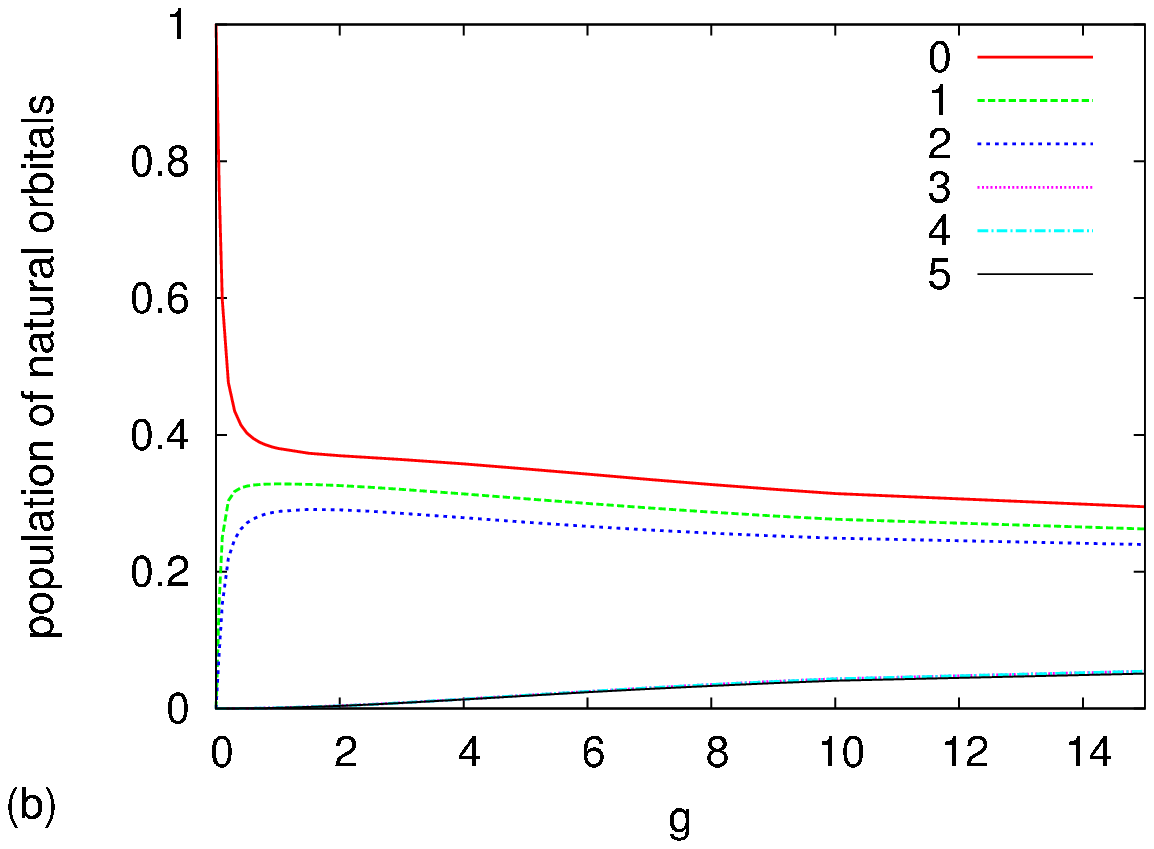}    
       \includegraphics[width=8.6 cm,height=4.3cm]{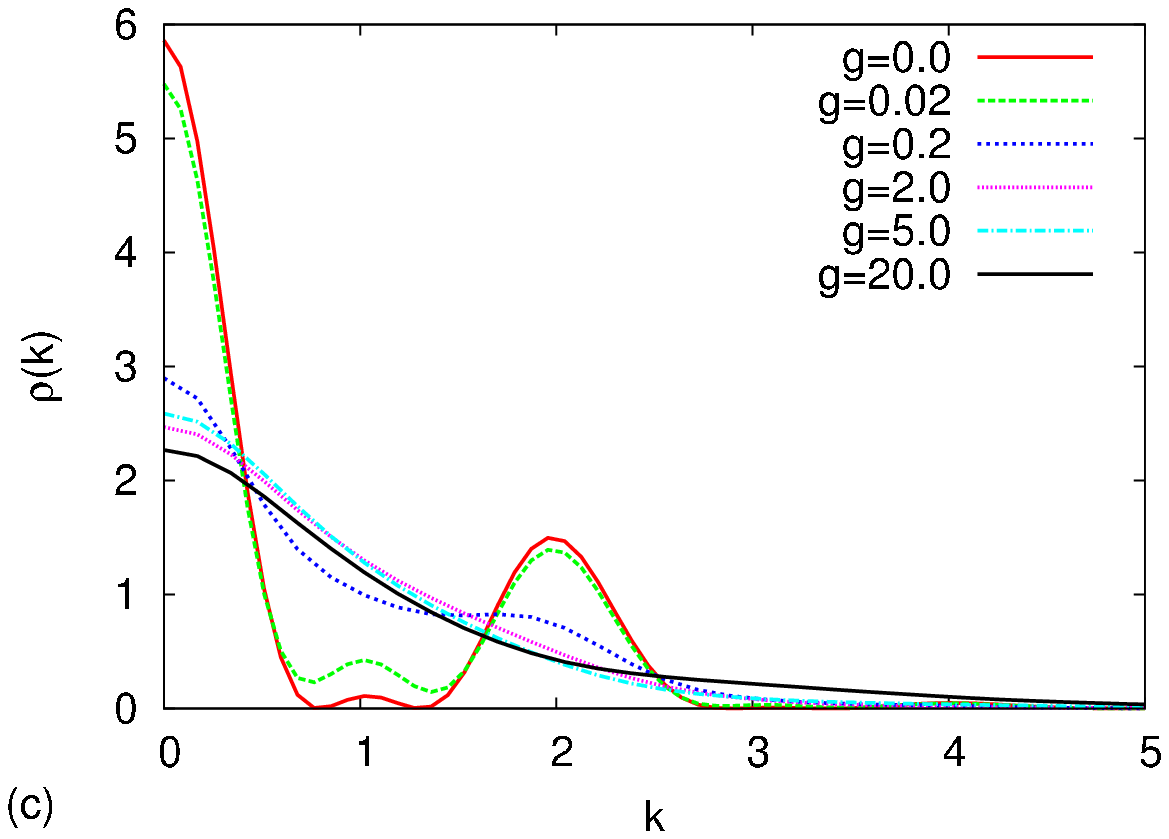}    
      \includegraphics[width=8.6 cm,height=4.3cm]{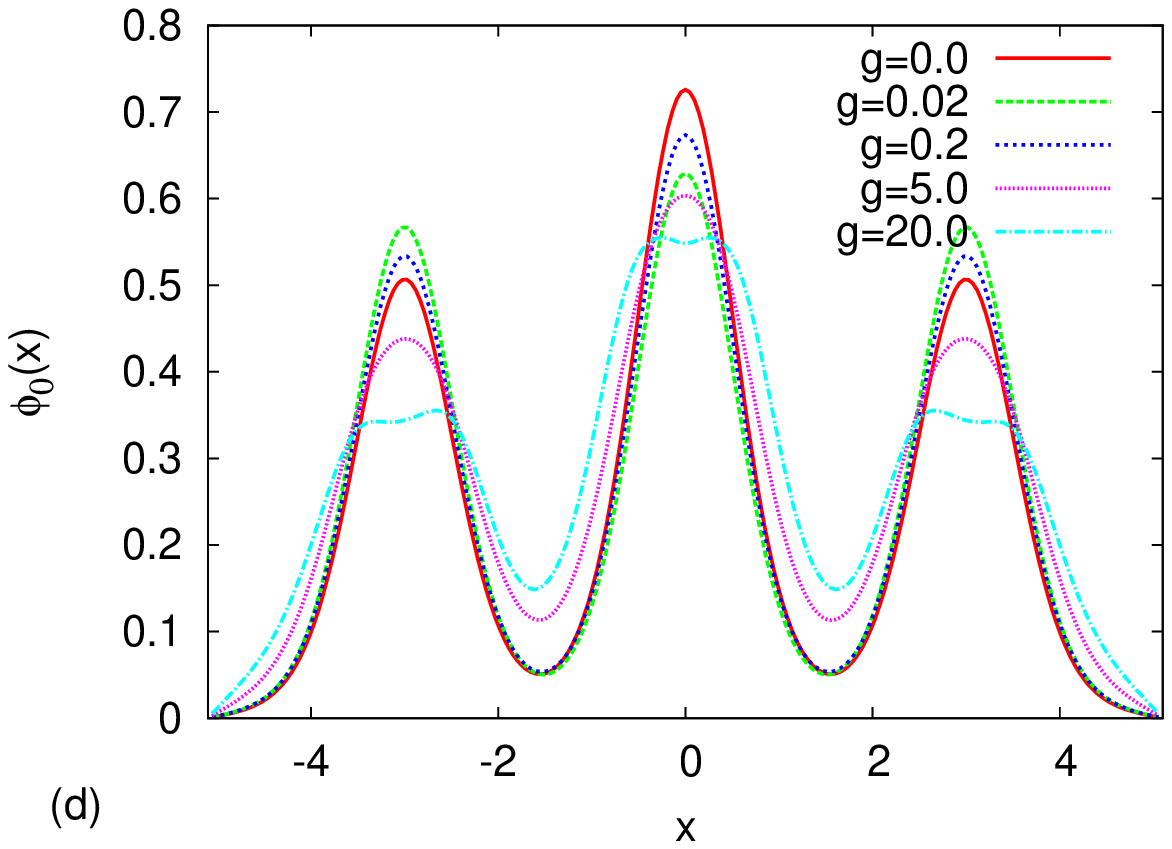} 

\caption{(Colour online) (a) $\rho_1(x,x')$ for 3 wells and 6 particles in the fermionization limit $g=20.0$. 
(b) Population of the natural orbitals as a function of the interaction strength.
(c) Momentum distribution: shown are 4 different values of $g$: non interacting ($g=0.0$), weakly interacting ($g=0.02, 0.2, 0.5$), on site fermionization crossover ($g=5.0, 20.0$). 
(d) Profile of the lowest natural orbital for several values of the interaction.}
  \label{fig8}

\end{figure}

For $\rho_1(x,x')$ the main effects are similar to the case of $\nu=1$. 
We comment that for the fermionization limit, $\rho_1(x,x')$ (Fig.~\ref{fig8} (a), $g=20.0$) reflects the isolated maxima of its diagonal part $\rho_1(x,x)=\rho(x)$,
 while the off-diagonal shows slight short-range correlations because of the broadening of the on-site functions. 
In the momentum distribution (Fig.~\ref{fig8} (c)), the interference peaks (Bragg and central one) become lower already for the onset of interactions,
 ending up in a complete smoothening for strong interactions $g=5.0,20.0$. 
The TG profile of the density ($g=20.0$) differs from the localised state of the BHM ($g=2.0$)
 in particular by the fact that the high momentum tails are more pronounced in the former case. 
Note that from $g=2.0$ to $g=5.0$ there is an increase of the $k=0$ peak which can be attributed to the on-site broadening of the density (similar to the case of the harmonic trap,
 see \cite{zoellner06b}). 
The immediate lowering of the central peak for small interactions is attributed to the deeper lattice that we use here
 compared to the case $\nu=1$ in the previous subsection ($V_0=12.0= 6.2 E_R$ for $\nu=1$ and $V_0=7.0=15.4E_R$ for $\nu=2$ with $d=3.3$). 
The fragmentation process is enhanced and thus the coherence is directly destroyed by admixing higher orbitals. 
Indeed, in a heuristic single particle picture, the energy levels within one band come closer for a deeper lattice and thus the gap is bridged easily by the interactions. 
The instant approach of the populations of the contributing natural orbitals of the lowest band (see Fig.~\ref{fig8} (b) 0,1,2) accounts for this fact. 
The population of the natural orbitals reflects the band structure: orbitals 0-2 of the effective first band and 3-5 of the second band (being indistinguishable in Fig.~\ref{fig8} (b)). 
The former orbitals (0-2) have the dominant contribution and are modulated,  following quite well the evolution of the one-body density,
 including the on site interaction effects (see Fig.~\ref{fig8} (d)).

\subsection{General remarks and energy properties}

Let us generalise our findings before we move to the case of an incommensurate filling. 
The equal distribution of the density onto the sites and the subsequent loss of fluctuations shown here, are also predicted by the BHM. 
These processes, which happen within the lowest band (and with Wannier states almost unaltered), are enhanced for a deeper lattice. 
On-site interaction effects occurring only for $\nu>1$, i.e, density broadening and formation of maxima as well as correlation hole and fragmented patterns in the two-body density,
 only show up for strong interactions (beyond BHM) when a substantial population of particles is well localised in one site. 
This fermionization crossover also applies to any integer filling factor, with the on-site phenomena involving $\nu=2,3,4...$ particles per well. 
The off-diagonal one-body correlations increase for increasing but weak interactions with the particles dominantly occupying the first orbital which flattens initially. 
For stronger interactions the fragmentation of particles destroys the coherence and the high visibility of the peaks in the momentum distribution is washed out to a smooth profile. 
For two particles per site the high momentum tails show differences between the 'BHM insulator' and the Tonks limit.

\begin{figure}

      \includegraphics[width=4.2 cm,height=4.2cm]{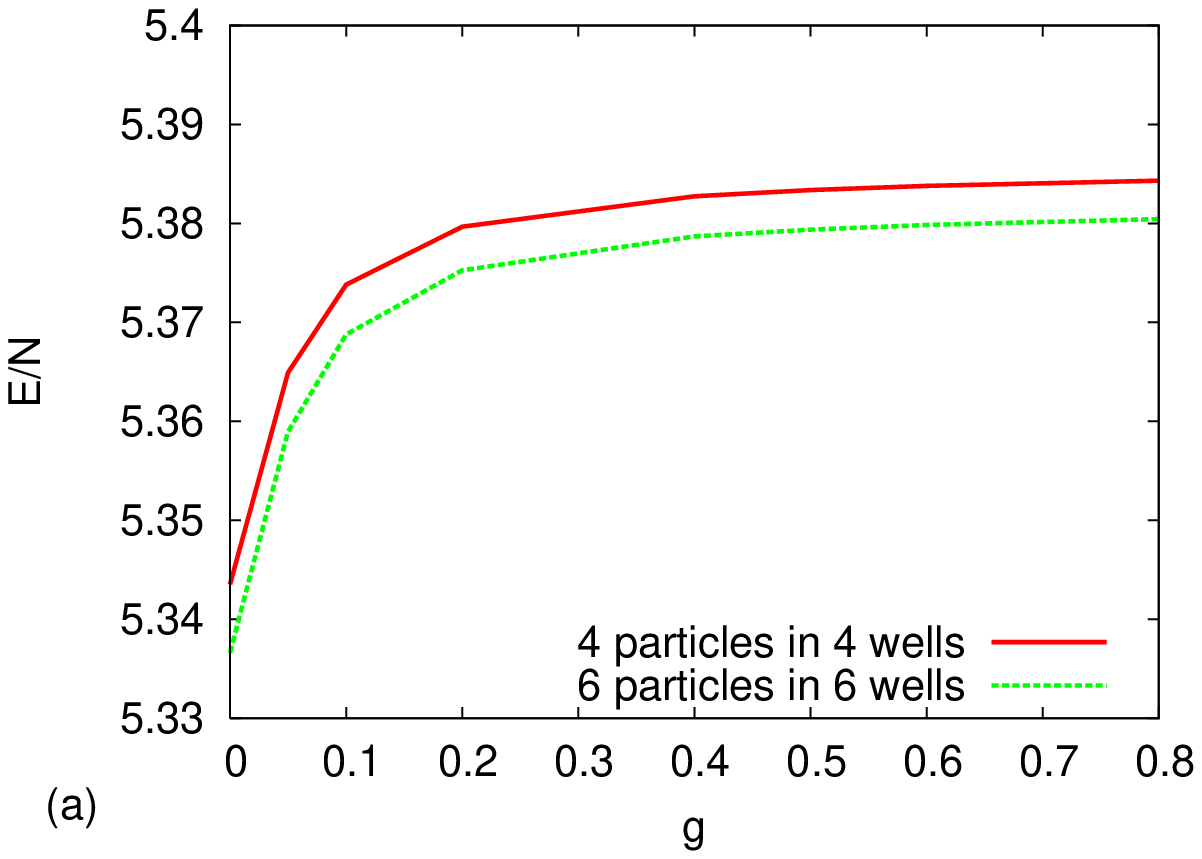} 
      \includegraphics[width=4.2 cm,height=4.2cm]{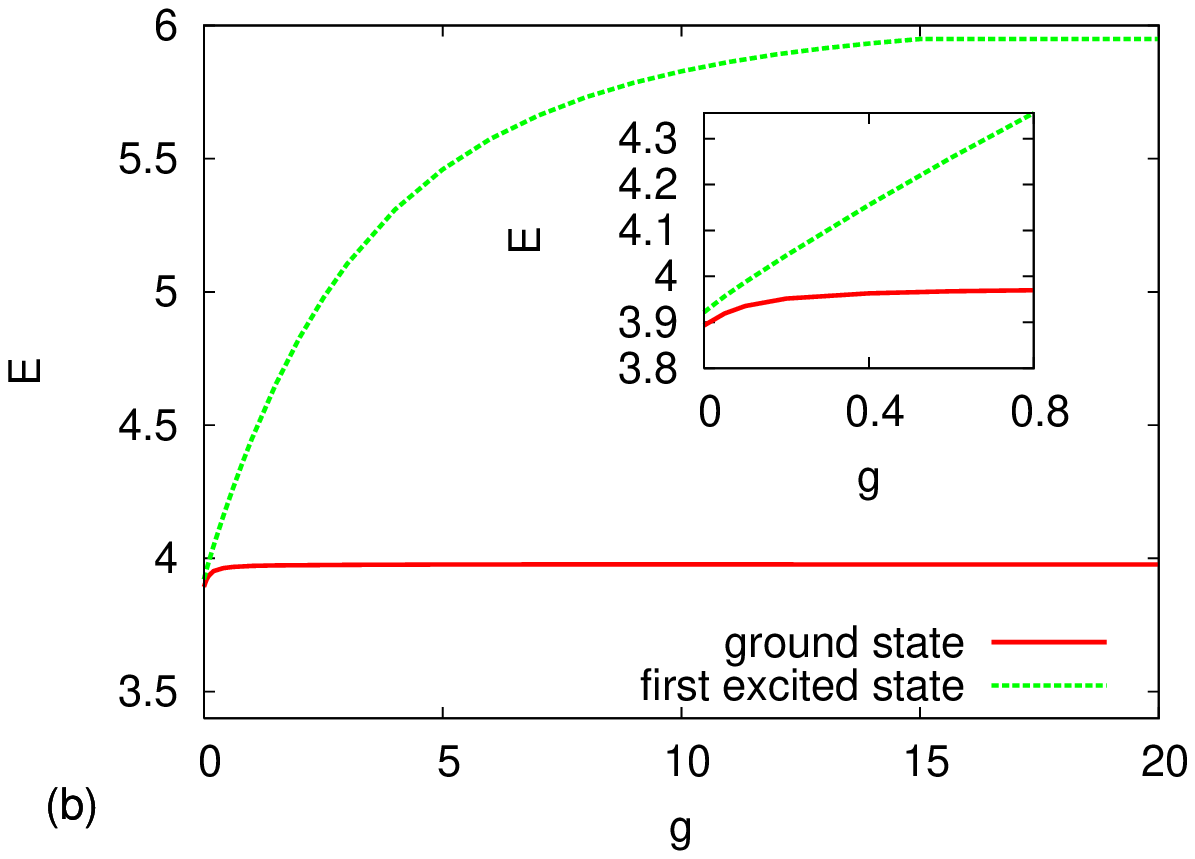} 

\caption{(Colour online) (a) Ground state energies per particle as the interaction increases for commensurate filling $\nu=1$ for the cases  4 particles in 4 wells and 6 particles in 6 wells. (b) Two first eigenstates of the spectrum of 3 particles in 3 wells as the interaction strength increases. (inset) weak interactions}
  \label{fig9}

\end{figure}

A general comment about the behaviour of the energy is in order here. 
The ground state energy increases with $g$ and saturates for $g \to \infty$ to the corresponding fermionic one, as we have laid out in Sec. II C. 
One particular aspect of this crossover is the response of the energy when we switch on the interactions.  
Close to $g=0$ the slope for the energy is approximately
\begin{equation}
\frac{dE}{dg}\Big\vert_{g=0} = \frac{N(N-1)}{2} \int |\varphi_0(x)| ^4dx.
\end{equation}
For hard-wall boundary conditions we know from Eq.~(2) that with increasing the number of wells (the size of the system),
 the population in the center of the lattice also increases if we keep the filling factor constant. 
Thus one would expect that the repulsive interaction affects larger systems more strongly due to the existence of areas with higher density. 
On the other hand, the integral of the single-particle ground state wave function to the fourth power representing the interaction term in the Eq. 5 above,
 is lower for a potential with more wells as the delocalisation of the state increases. 
The above effects cancel out resulting in an almost size-independent evolution of the ground state energy per particle as long as $N=W$ [see Fig.~\ref{fig9}(a)]. 
Of course, for increasing filling factors interaction effects manifest more strongly in the evolution of the energy as the density plays the dominant role.
 
A second aspect is the energy gap between the ground and the first excited state. 
Let us point out that even without interactions, we have here an energy gap between the two states of the order of $J$ [see Fig.~\ref{fig9}(b)];
 a continuous band structure which results in a gapless spectrum in the SF regime arises only in the limit of an infinite lattice. 
Moreover, since the Bose-Fermi map holds also for excited states, we can compute exactly the gap for $g \to \infty$,
 which is of course the interband gap of the single particle spectrum. 
In the BHM regime (see Fig.~\ref{fig9}(b) inset), the gap is of the order of $U$, but in general, as considered here,
 it is evolving continuously and occurs not as a sudden transition as it does in the macroscopic case.

The last but important comment refers to the role of the kinetic energy, which is related to the parameter $J$ and this, in turn, to the lattice depth. 
We have already underlined its impact on the enhancement of localisation and fragmentation for a lower $J$. 
We performed  all the calculations assuming a sufficiently deep lattice, such that at least two single particle bands lie below the energy maxima of the barriers (see Fig.~\ref{fig1}). 
This choice validates our argumentation in terms of the tight-binding approximation and also it ensures validity of the BHM for small interactions. 
In the other limit of a shallow lattice, a hydrodynamic approach in the framework of the sine-Gordon model has been employed \cite{buechler03}. 
They show that, for very strong interactions in 1D, an arbitrarily small perturbative lattice potential is enough to result in an insulating phase for commensurate filling. 
However, the latter discussion is based on the thermodynamic limit. 
For our system, if we have particles delocalised above the barriers,
 then the on-site few-body effects smoothen out, and there is an interaction induced broadening leading to a filling of the space between the wells. 
In general as the total energy increases the particles come energetically closer to the continuum,
 where the barriers thin out and thus there is an enhanced penetration into the barrier (flattened on-site functions) which effectively results in larger fluctuations. 
A hint towards this effect is already evident in the case of  $\nu=2$ in the one body density [Fig.~\ref{fig6} (a)]
 where for strong interactions $g=5.0,20.0$ we see a slightly higher density in the inter-well space. 
Accordingly the fluctuations [Fig.~\ref{fig6} (b)] show a slope to slightly higher values .

\section{Incommensurate Filling}

Incommensurate filling is more susceptible to the exact number of particles compared to the number of wells, with one main feature: there is always a delocalised fraction of particles. 
We consider essentially two cases of $\nu$ non-integer: $\nu<1$ where on-site interaction effects do not manifest due to low population
 and $\nu>1$ which for strong interactions can be interpreted as a fraction 
 $N~\mathrm{mod}~{W}$ of extra delocalised particles sitting on a commensurate background of localised particles \cite{pupillo06}.

\subsection{Filling factor $\nu<1$}

\begin{figure}

      \includegraphics[width=8.6 cm,height=4.3cm]{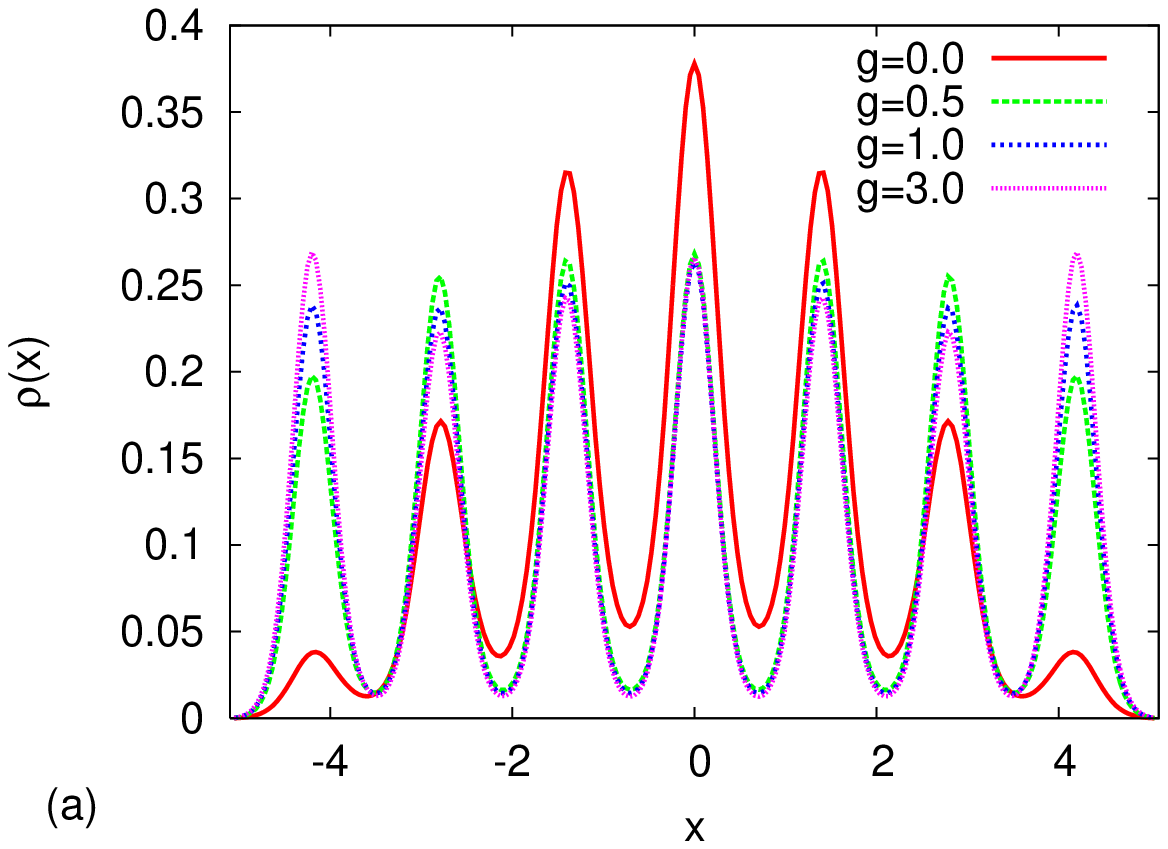} 
      \includegraphics[width=8.6 cm,height=4.3cm]{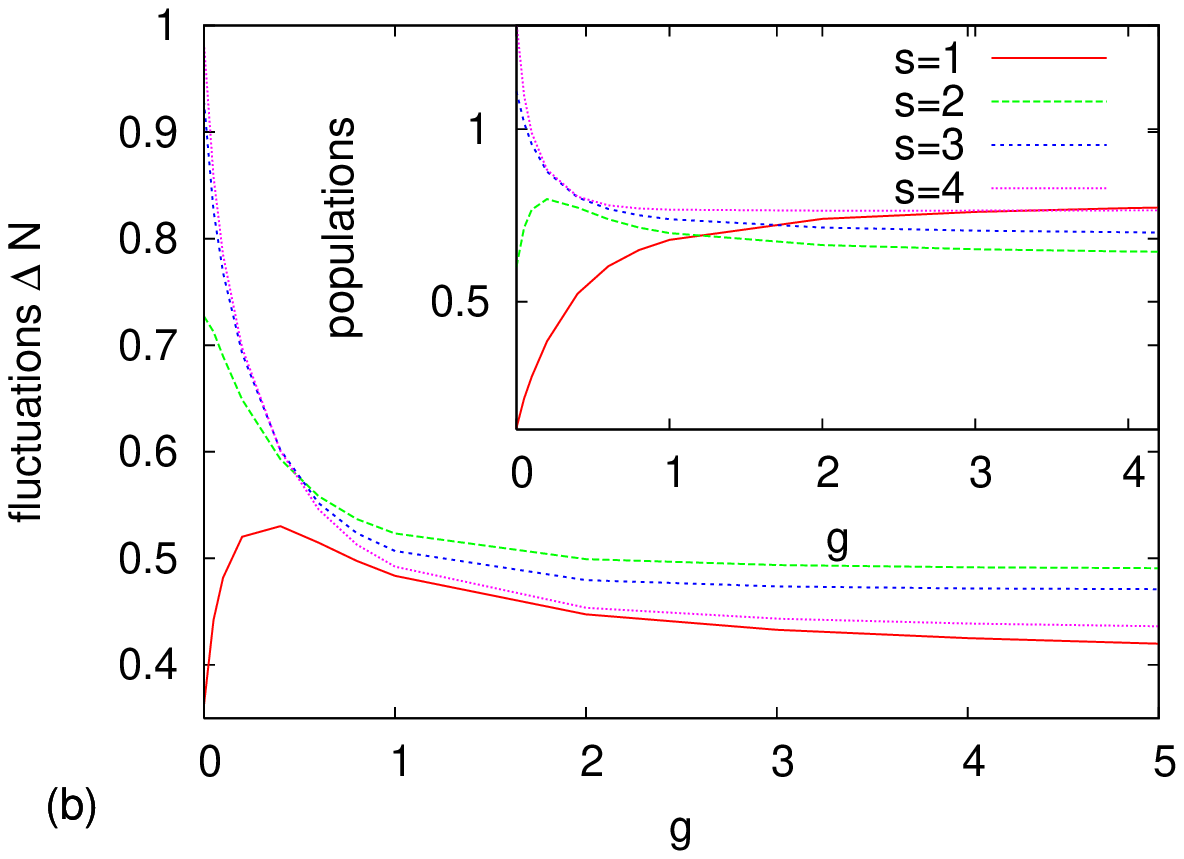} 

\caption{(Colour online) (a) $\rho(x)$ for 5 particles and 7 wells. Shown are 4 different values of $g$: non interacting ($g=0.0$), weakly interacting ($g=0.5, g=1.0$) and fermionized limit ($g=3.0$). 
(b) Particle number fluctuations as g increases for the sites $s=1,2,3,4$. (inset) On-site population.}
  \label{fig10}
\end{figure}

The main concern here is how the particles distribute over the lattice as the repulsion increases. 
In the weak interaction regime, the repulsive forces drive the particles away from the highly populated center of the potential,
 (see for example 5 particles in 7 wells  in Fig.~\ref{fig10} (a)). 
In this case the one-particle density does not tend to an equal site occupation, but stays even for strong interactions asymmetric ($g=3.0$). 
The exact number of particles $N$ and wells $W$ determines the fermionized distribution of the density:
\begin{equation}
\rho(x)_{\mathrm{Fermi}} \propto \sum_s^W \sum^N_q |\sin \left(\frac{s q \pi}{W+1} \right)|^2 |w_s(x)|^2
\end{equation}
where $q$ includes only lowest-band states since $N<W$. 
Hence, the BHM is valid for the whole range of interactions. 
We can understand the final profile also in a hole-excitation picture where starting from the the MI in the commensurate case $N=W$
 we annihilate $W-N$ particles $\Psi= \sum_{\alpha=1}^{N-W} a_{\alpha}|\mathrm{MI}_{N=W} \rangle$. 
While in the case of $\nu=1$ the addition of $N=W$ coefficients in Eq. 6 leads to equal site occupation,
 here we can have imbalances and oscillations of the density depending on how many orbitals contribute according to the numbers $W$ and $N$
 (Fig.~\ref{fig10}(a) $g=3.0$, Fig.~\ref{fig10}(b) inset). 
Triggered by the interaction, the 'transfer' of particles from the middle of the potential to the outer positions passes through the intermediate wells which gain
 and lose population (Fig.~\ref{fig10}(b) inset $s=2$). 
The number fluctuations in Fig.~\ref{fig10} (b) saturate to a rather high value because of the incommensurate filling $\nu<1$ which allows only delocalised phases. 
The fluctuations are greater in the wells with less population corresponding to 'holes' (Fig.~\ref{fig10} (b) compare $s=2,3$ with $s=1,4$). 

\begin{figure}

      \includegraphics[width=3.6 cm,height=3.6cm]{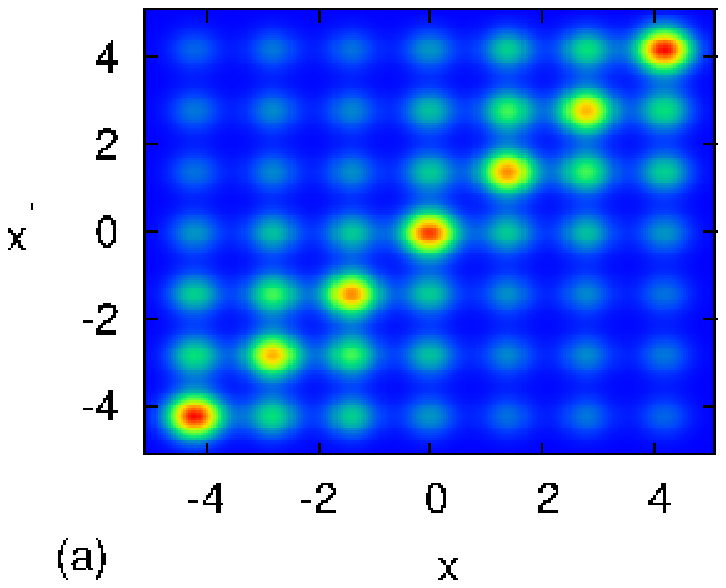} 
      \includegraphics[width=4.3 cm,height=4.0cm]{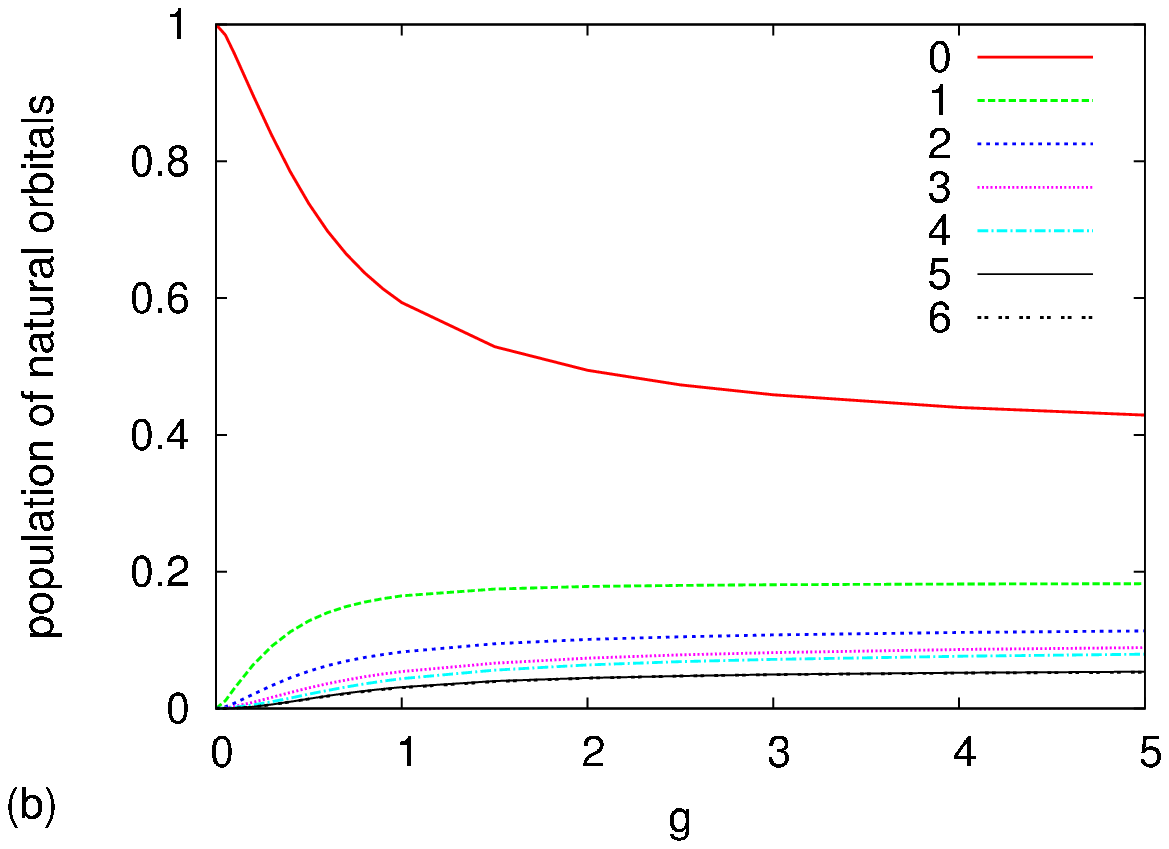} 
      \includegraphics[width=8.6 cm,height=4.3cm]{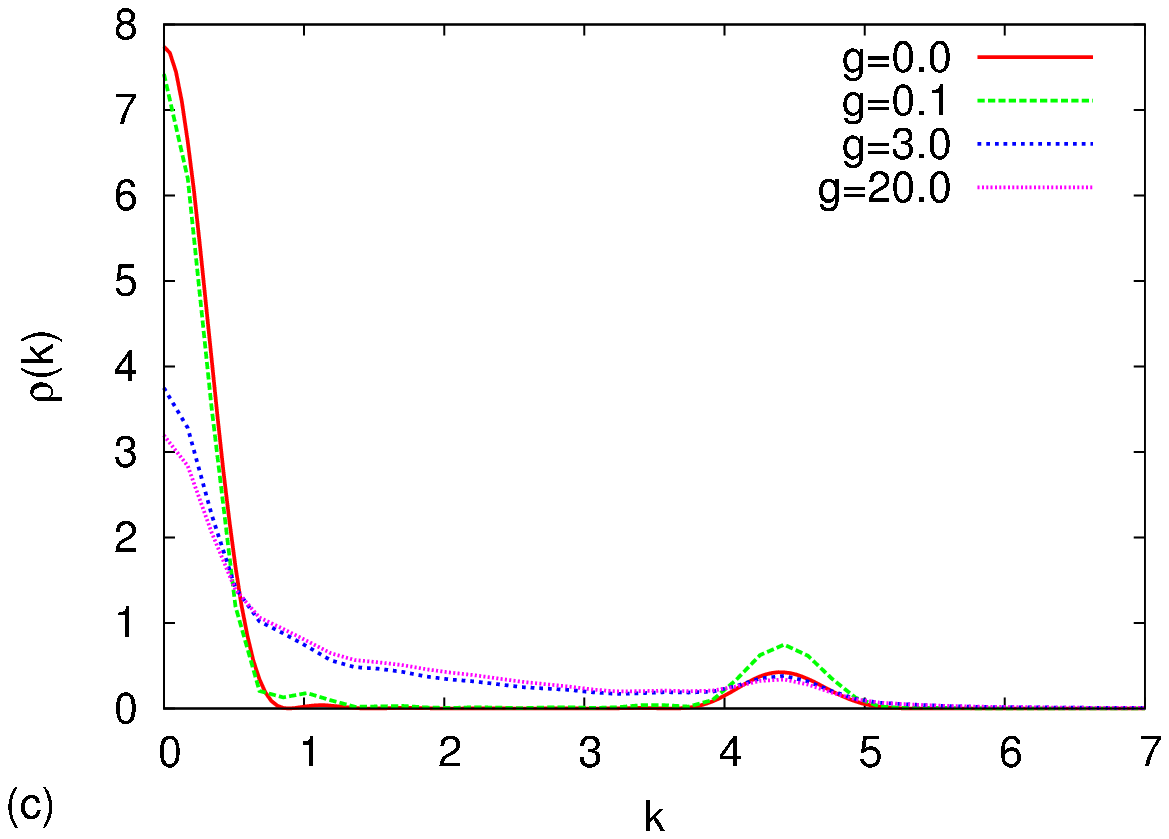}

\caption{(Colour online) (a) $\rho_1(x,x')$ for 7 wells and 5 particles in the fermionization limit. 
(b)   Population of the natural orbitals as a function of $g$.
(c) Momentum distribution for 5 different values of $g$: non interacting ($g=0.0$), weakly interacting ($g=0.1$), fermionization limit ($g=3.0, 20.0$).}
  \label{fig11}
 
\end{figure}

A main difference between commensurate and incommensurate filling is that in the latter case the 'coherence',
 or better the off-diagonal part of $\rho_1(x,x')$, cannot vanish completely since the particles remain in fact delocalised. 
The remaining one-body correlations in the fermionization limit  (Fig.~\ref{fig11} (a)) are concentrated mostly close to the diagonal, i.e., between neighbouring lattice sites. 
It is interesting though that this short range coherence is not equally distributed on all close to the diagonal spots. 
For example, in the case of 5 particles in 7 wells,
 there is a quite well localised particle in the central site as the vanishing off-diagonal terms in the center of the figure indicate. 
The distribution of short-range correlations is again related to the commensurability of the setup, and can be understood in the following way:
 divide the real space of the multi-well potential into N equal intervals and put each particle in the middle of one interval;
 then those which lie closer to the middle of a site are also better localised than those who lie close to the inter-well barriers affecting the one-body correlations accordingly,
 i.e., when there is more localisation the correlations die out.

The attempted localisation of the particles as the repulsion increases distorts to some extent the interference pattern in the momentum distribution (Fig.~\ref{fig11} (c)). 
In particular the central peak is lowered due to the partial loss of coherence but the peaked structure is not fully smeared out. 
We do not observe an increase of the central peak of the momentum for low interactions as in the case of $\nu=1$,
 even though we use a rather shallow lattice $V_0=10.0=4.1E_R$ with $d=1.42$). 
Although the first natural orbital is broadened  and dominant [Fig.~\ref{fig11} (b)],
 the density of 'condensed'  particles $\frac{n_0 N}{W}$ is not as high in this case of incommensurate filling $\nu<1$ as for the case $\nu=1$
 and this explains the instant lowering of the $k=0$ peak. 
As expected $W$ natural orbitals contribute substantially (see Fig.~\ref{fig11} (b)) as for $\nu=1$,
 but the contribution of each orbital here, differs throughout the fermionization crossover.

\subsection{Incommensurate Filling $\nu>1$}

The incommensurate $\nu>1$ case combines localisation delocalisation as well as on-site interaction effects. 
It is instructive to keep in mind the properties of the corresponding $\nu^<=\frac{N\mathrm{mod}W}{W}$ filling case,
 which refers only to the 'extra' $N\mathrm{mod}W$ particles, thereby pointing out similarities and differences which allow to identify the effect of the 'background'. 
For all the following cases the length unit is $L/9$ and $d=2.2$ (four wells). 
The lattice depth is chosen $V_0=20.0=19.6E_R$ such that the particles are confined energetically below the continuum. 

\subsubsection{One extra particle on a unit filling background}

\begin{figure}

       \includegraphics[width=8.6 cm,height=4.3 cm]{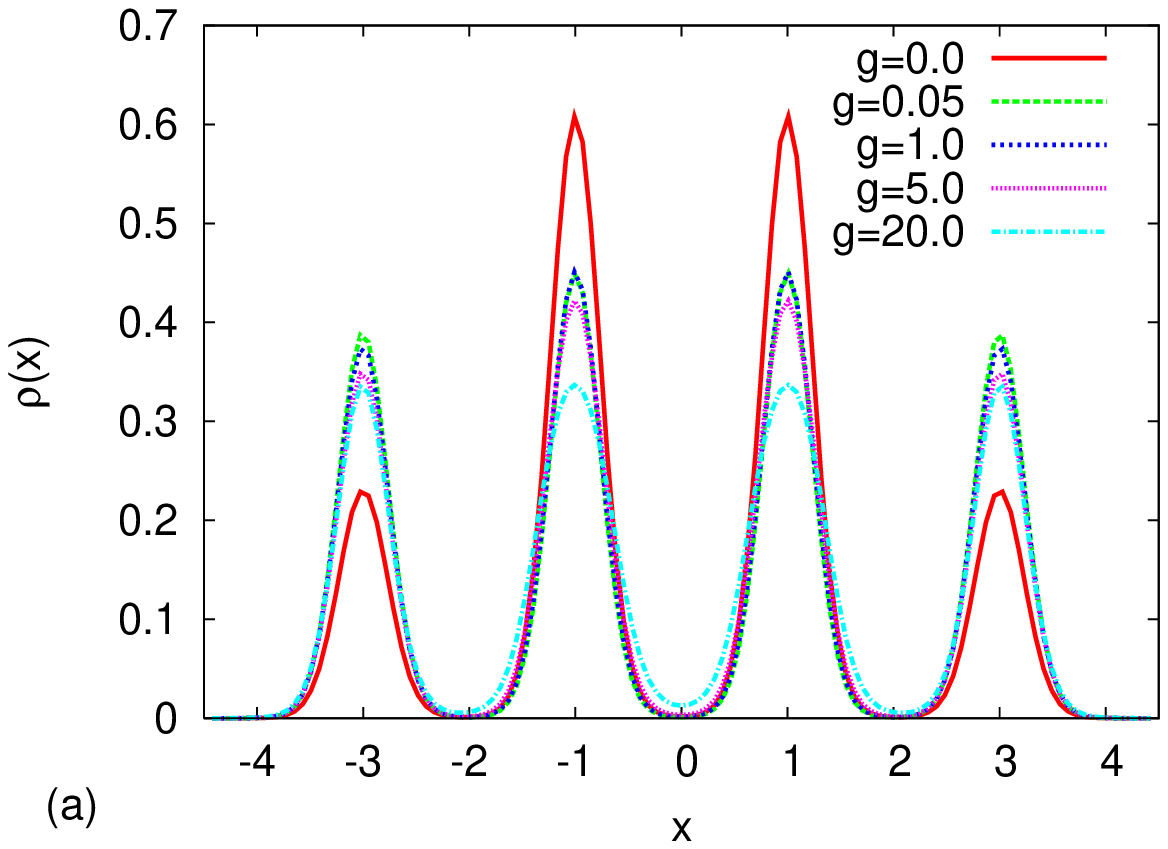} 
       \includegraphics[width=8.6 cm,height=4.3 cm]{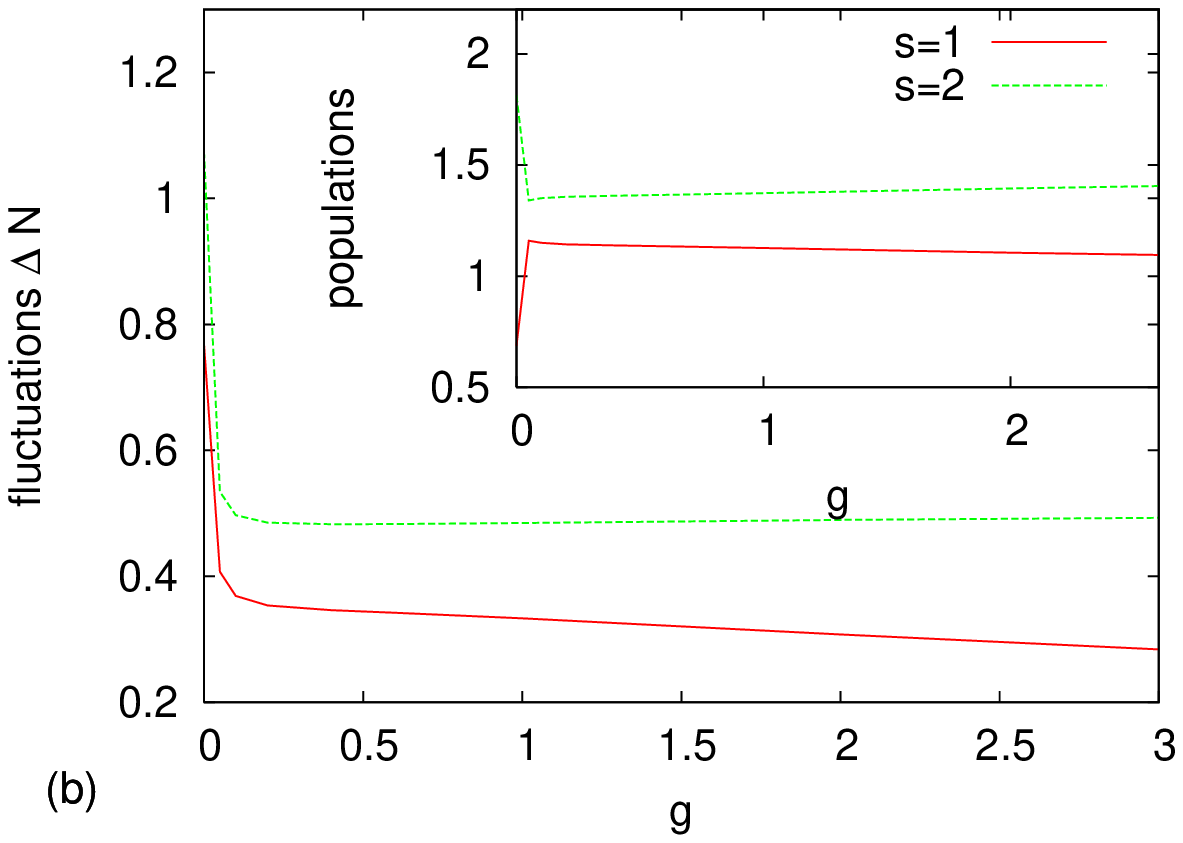}

\caption{(Colour online) (a) $\rho(x)$ for 5 particles and 4 wells. Shown are 4 different values of $g$: non interacting ($g=0.0$), weakly interacting ($g=0.05, 1.0, 5.0$), fermionization limit ($g=20.0$). 
(b) Particle number fluctuations as $g$ increases for the sites $s=1,2$. (inset) On-site populations. }
  \label{fig12}

\end{figure}

We begin with the simplest case of one extra particle added to the unit filling (here 5 particles in 4 wells). 
For weak interactions the $g=0$ nonuniform occupation of sites tends to become uniform [Fig.~\ref{fig12}(a) $g=0.05$]. 
For a slightly higher interaction ($g=1.0$) strength though, there is an interesting revival of the tendency to predominantly occupy the middle wells. 
To understand this, we need to go beyond the lowest band and BHM analysis (since $\nu>1$) and consider contributions of higher bands. 
The higher band states possess a similar distribution with respect to the different wells as the lowest band (see Eq. 2 ). 
Thus the energetically lowest level of the excited band, which has a dominant population in the middle (see eg. Fig.~\ref{fig1} for the triple well),
 when contributing, results in repopulation of the center. 
Additionally as the interaction increases, the total energy of the particles becomes higher
 and thus they approach energetically the top of the barriers of the potential close to the continuum (see also the case $\nu=2$). 
This enforces the hopping term since higher bands with larger coupling $J^{1}>J^{0}$ contribute (compare intra-band splittings in Fig.~\ref{fig1})
 and as a consequence the kinetic energy term redirects the particles to the center. 
This intuitive picture of contributions from higher energy states that we have drawn here for the repopulation of the middle,
 is also consistent with the treatment of the Tonks limit via the Bose-Fermi map. 
According to the theorem, the extra particle lies exactly on the energetically lowest one particle level of the excited band, which possesses also higher contributions in the middle,
 while the other particles completely occupy the lowest band states forming a MI background of one particle localised per well. 
This results in a slight broadening especially of the central peaks of the one-body density (Fig.~\ref{fig12}(a) $g=20.0$), a standard on-site interaction effect for $\nu>1$,
 which we encountered also in the previous section for $\nu=2$. 
Let us note that the situation is quite different for a sufficiently shallow lattice not considered here:
 the extra particle would go closer or even above the barriers and thus would fill the inter-well space,
 distributing smoothly over the potential and resulting in strongly reduced on-site effects \cite {zoellner06a}. 
A reference model to understand incommensurate filling $\nu>1$, in qualitative agreement with our results, was given in \cite{pupillo06}:
 the particles occupy different horizontal 'layers', each one on top of the other, all having commensurate MI states and only the highest one being incommensurate
 (with $N \mathrm{mod} W$ particles) and delocalised.

In the evolution of the populations with increasing $g$ we can verify the density variation in the center wells (see Fig.~\ref{fig12}(b) inset). 
In the strongly interacting regime, the populations remain quite similar due to the background of localised particles and they differ
 only because of the non-uniform distribution of the extra particle in the first level of the excited band. 
The particle-number fluctuations  (Fig.~\ref{fig12}(b)) remain quite large, since the extra delocalised particle does not allow for a perfect insulator phase. 
Nevertheless they are substantially diminished compared to the corresponding $\nu^<=1/4$ single particle case  because of the localisation of the background. 

\begin{figure}

      \includegraphics[width=3.6 cm,height=3.6 cm]{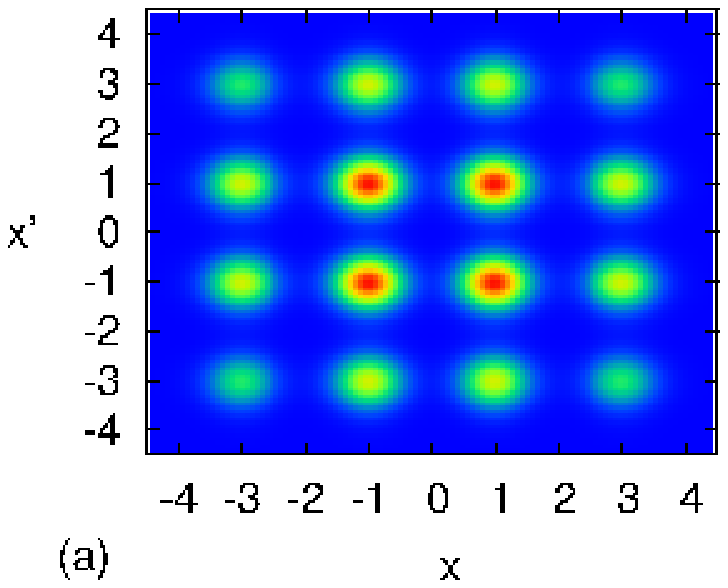} 
      \includegraphics[width=3.6 cm,height=3.6 cm]{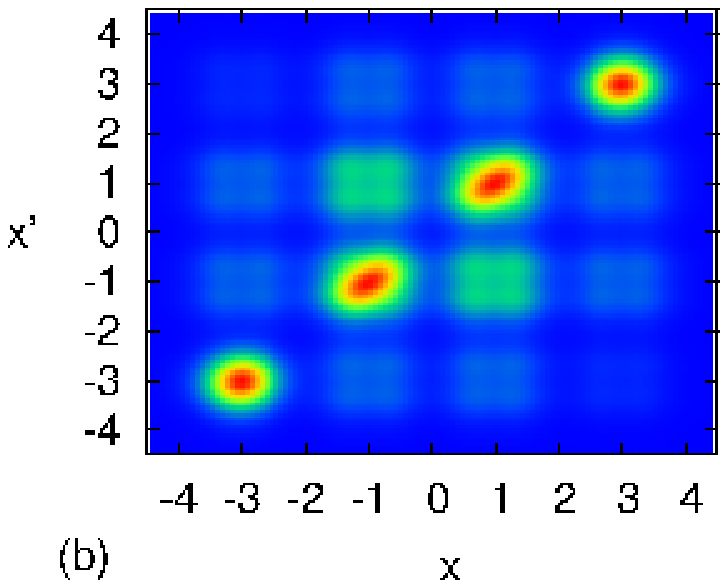} 
      \includegraphics[width=8.6 cm,height=4.3 cm]{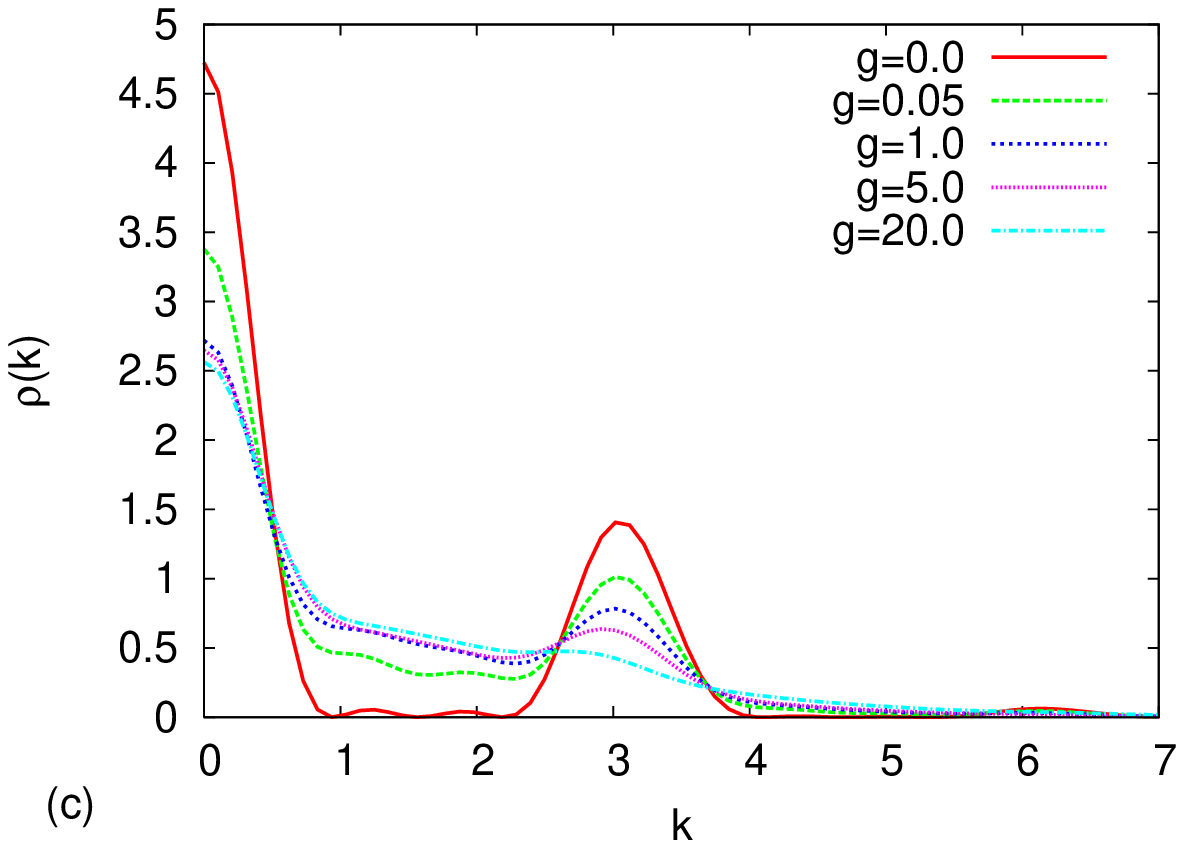}

\caption{(Colour online) (a) One-particle density matrix $\rho(x,x')$ for 5 particles in 4 wells in the non-interacting $g=0.0$ and (b) in the fermionization limit $g=20.0$. (c)  Momentum distribution. Shown are 5 different values of $g$: ($g=0.0,0.05,1.0,5.0,20.0$). (d) Population of the natural orbitals as a function of the interaction strength.}  
  \label{fig13}

\end{figure}

The non-local properties confirm the incomplete localisation in this case. 
In the one body density matrix $\rho_1(x,x')$ the remaining coherence in the strongly interacting limit (Fig.~\ref{fig13} (b)), is concentrated close to the diagonal. 
Due to the localised background the long range off-diagonal terms almost disappear, compared to the $\nu^<=1/4$ case of a single particle (Fig.~\ref{fig13} (a)). 
The off-diagonal humps, which are mainly visible in the center, reflect a widened pattern (Fig.~\ref{fig13} (b)). 
In the momentum distribution (Fig.~\ref{fig13}(c)) the central peak and the Bragg peaks are from very weak interactions ($g=0.05,1.0$)
 lowered because of the large depth of the potential used in this case ($V_0=20.0=19.6E_R$). 
Small peaks still persist, revealing incomplete localisation due to the extra particle (Fig.~\ref{fig13}(c) for $g=20.0$).

\subsubsection{A repulsive pair of particles on a localised background}

\begin{figure}
 
        \includegraphics[width=8.6 cm,height=4.3cm]{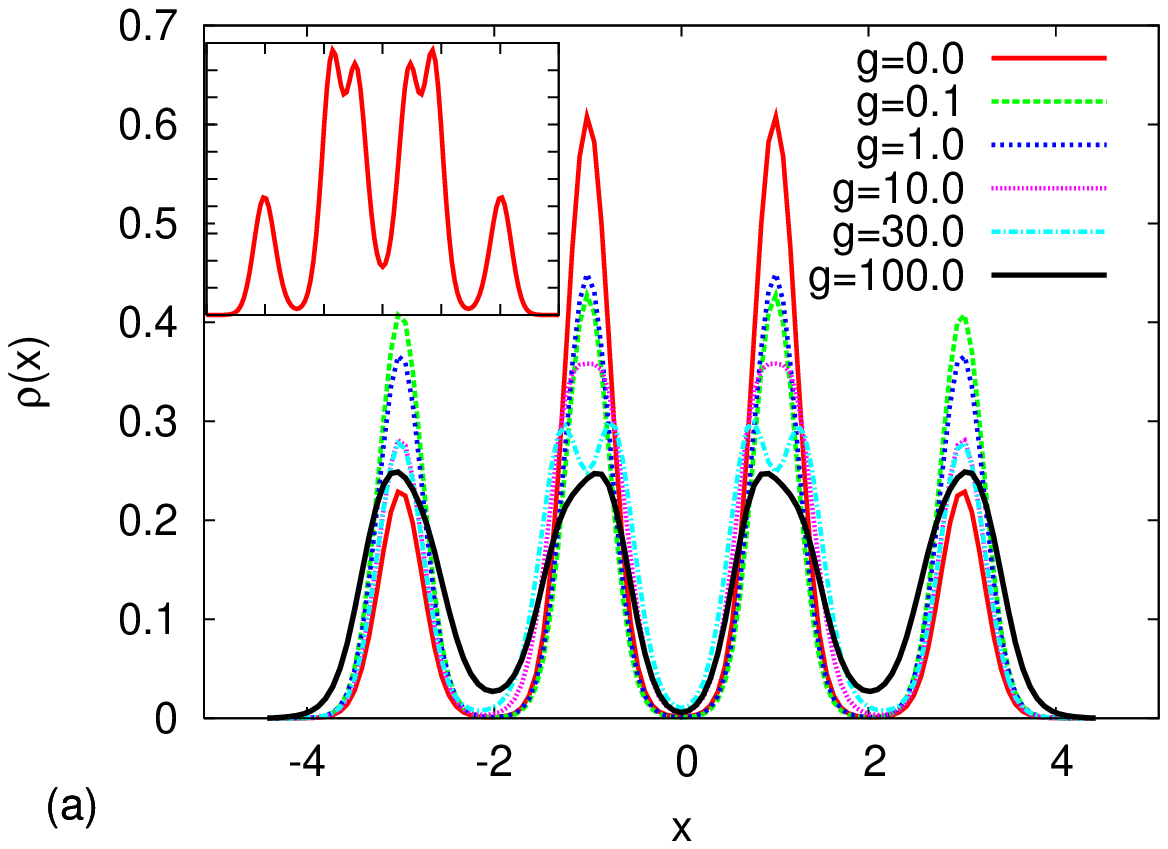} 
	\includegraphics[width=4.2 cm,height=4.2cm]{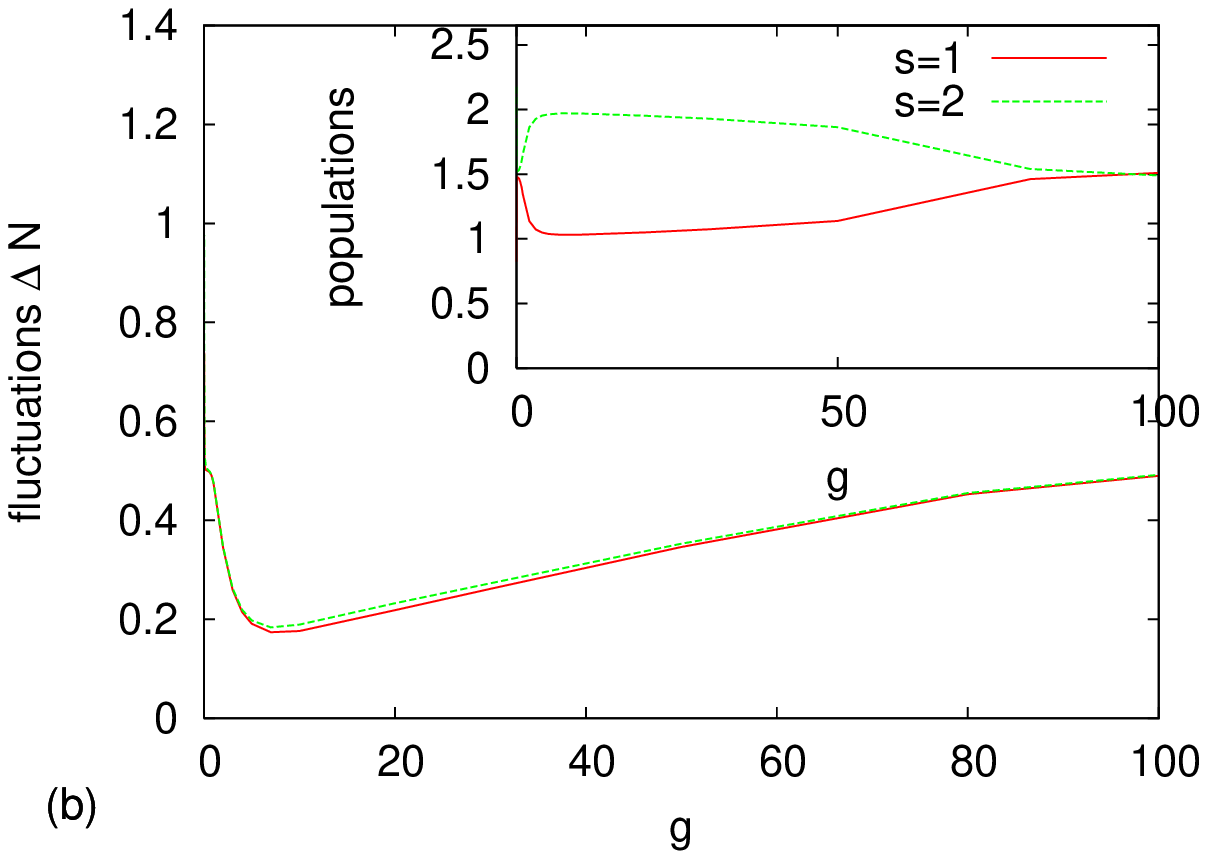}  
	\includegraphics[width=4.2 cm,height=4.2cm]{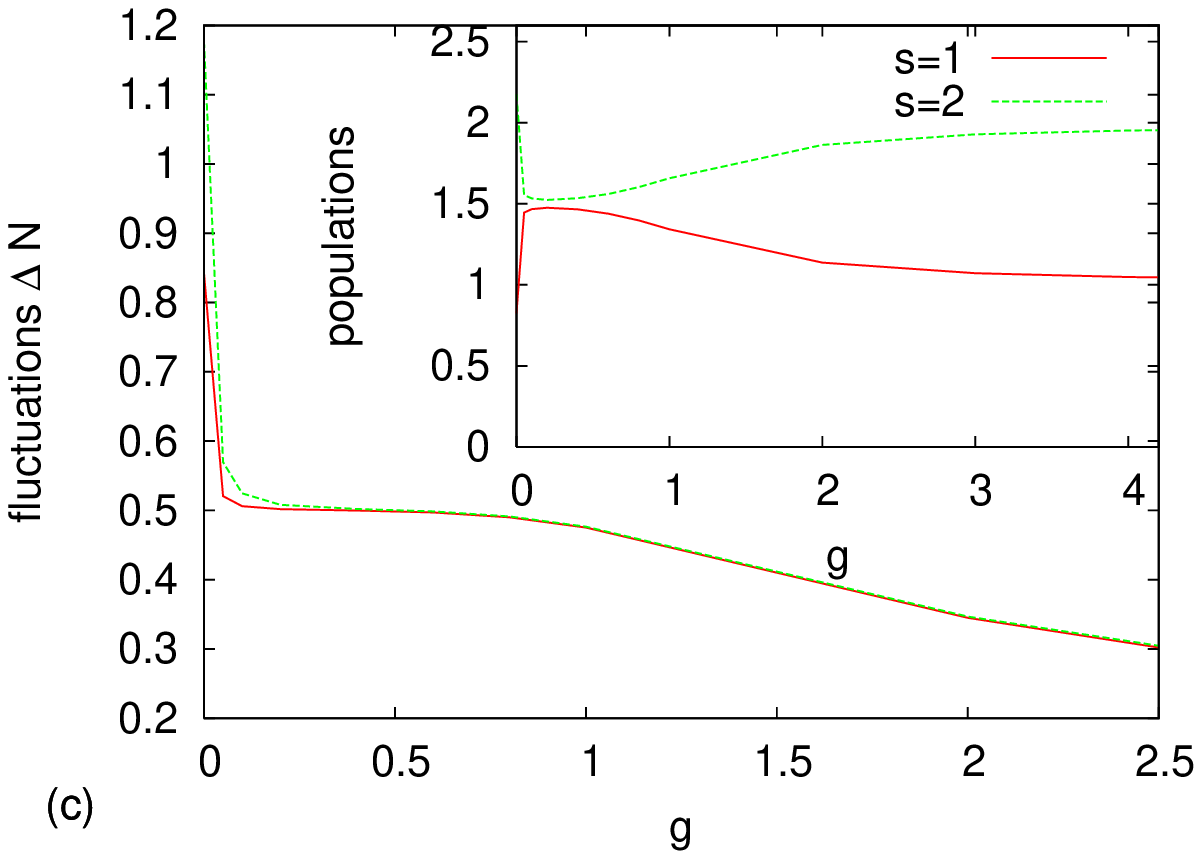} 
        \includegraphics[width=8.6 cm,height=4.3cm]{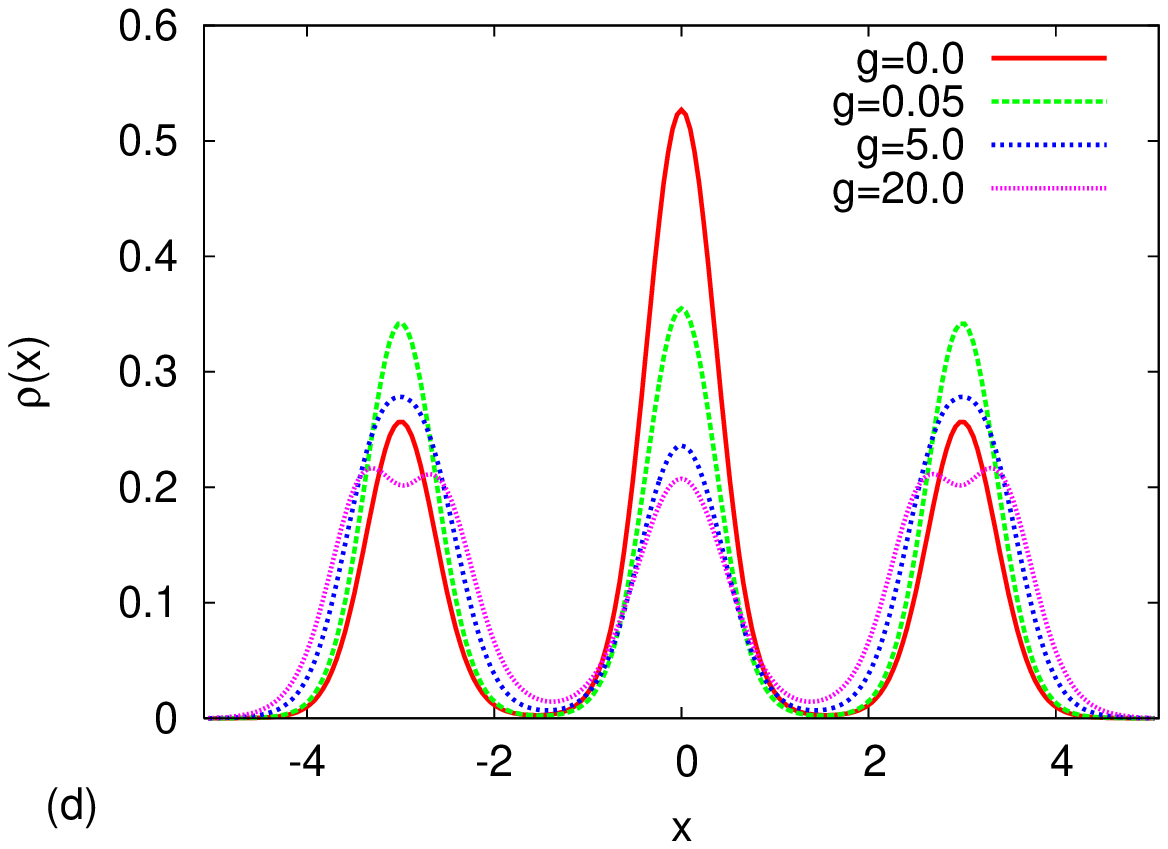}
\caption{(Colour online) (a) One body-density for 6 particles and 4 wells. 
Shown are 6 different values of $g$: non interacting ($g=0.0$), weakly interacting ($g=0.1, 1.0$), strongly interacting ($g=10.0, 30.0$) and fermionization limit ($g=10.0$). (inset) The case of a superimposed harmonic trap with strong interactions $g=30.0$ for 5 particles.  (b) Particle number fluctuations as a function of $g$ for $s=1,2$. (inset) On-site populations. (c) Same plot for the weak interaction regime. (d)  One-body density for 5 particles in 3 wells for $g=0.0, 0.05, 5.0, 20.0$. }
  \label{fig14}

\end{figure}

Choosing the number of particles and wells at will, one encounters many different incommensurate cases and corresponding effects. 
We focus here on the case of incommensurate filling with two extra particles, which exhibits a distinct behaviour:
 the extra pair of particles feels the interaction with the background particles, additionally to the intra-pair repulsive forces. 
From the above discussion (sec IV. B.1), we recall tendencies to first equalise and then repopulate the center wells with increasing $g$. 
Indeed this happens also in the low-interaction regime for the case of 6 particles in 4 wells examined here (see Fig.~\ref{fig14}(a) $g=0.1,1.0$, Fig.~\ref{fig14}(c)). 
It even results in a broadening of the central peaks and wiggles in the one-body density for higher interaction strengths (Fig.~\ref{fig14}(a) $g=10.0, 30.0$). 
To realise the peculiarity of this effect one has to consider the fermionization limit which exhibits only a slight broadening but no wiggles at all;
 rather the particles are distributed uniformly (Fig.~\ref{fig14}(a) and Fig.~\ref{fig14} (b) for $g=100.0$) in accordance to the Bose-Fermi map
 which predicts equal contribution from the two first levels of the excited band. 
However, for strong but finite interactions the extra particles concentrate more in the center resulting in an effectively higher local filling $\nu_{s=2,3}\approx 2$;
 only for $g>30$ the repulsion is strong enough to drive the bosons to the outer wells. 

In Fig.~\ref{fig14}(c) we observe that for weak to intermediate interactions ($g \approx 0.05-0.8$), the populations and the fluctuations are almost constant. 
Beyond this regime (close to $g=0.8$), the population of the center wells ($s=2$) increases and approaches the value for two particles in these sites $\nu_{s=2,3}\approx 2$,
 while the fluctuations are again strongly reduced. 
This can be understood as a localisation behaviour ('Mott-like phase') in a central 'domain' of the potential
 in analogy with similar situations appearing for an optical lattice with a superimposed {\it harmonic} confinement
 \cite{batrouni02,rey05,gerbier06,campbell06,rigol09,roux08,luehmann09,scalettar91,damski03}. 
In the latter case, the harmonic potential increases the on-site energy as we go to the outer wells and thus sets an 'energetic obstacle' for the particles to occupy them. 
Therefore, they prefer to localise in the center and form Mott domains (or shells), possibly surrounded by a superfluid layer of delocalised particles. 
As an illustration of this effect in our few-body setup, we present the case of 5 particles in a 1D lattice with a superimposed {\it harmonic}  trap (Fig.~\ref{fig14}(a) inset). 
For strong interactions ($g=30.0$ here) we have exactly two particles in each middle well and one particle divided into the two outer wells surrounding the 'Mott-shell'. 
We underline that our numerically exact method which goes beyond BHM brings a new light on the strongly interacting regime including the on-site interaction effects; 
here we point out the formation of wiggles in the one-body density inside the 'Mott-shell' of two particles per site. 
For our initial setup though there is no harmonic confinement and thus no 'energetic obstacles' between the wells; nevertheless the hard-wall boundary conditions on the edges,
 make it preferable for the particles to be in the center. 
We can thus comment that the finite size itself makes it possible for qualitatively different spatial regions to occur in incommensurate one-dimensional lattices. 
For stronger interactions, the fluctuations increase again (Fig.~\ref{fig14} (b)) as the particles occupy higher energy levels and delocalise further. 
The fermionization comes with equal population $\frac{N}{W}=1.5$ particles per well (Fig.~\ref{fig14} (b) inset).

Similar phenomena of course can happen for other cases of incommensurate filling $\nu>1$. 
Let us stress that the exact number of wells and particles is important for the possible effects. 
For example, in the case of 5 particles in 3 wells (Fig.~\ref{fig14} (d)), the repulsively interacting extra pair of particles is prohibited from occupying the middle well,
 since it would require a very undesirable triple occupation of a single site with strongly repulsive particles. 
Thus the revival of occupation in the center is avoided, and each of the two extra particles is mostly located in one outer well,
 indicated by the corresponding wiggles of the one-body density. 

\begin{figure}
  \begin{center}
    \begin{tabular}{cc}
      \includegraphics[width=4.0 cm,height=4.0 cm]{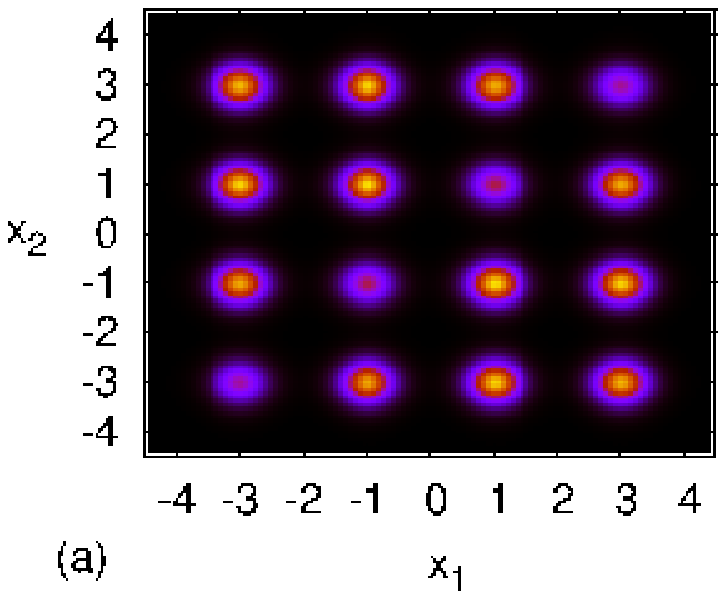} &
      \includegraphics[width=4.0 cm,height=4.0 cm]{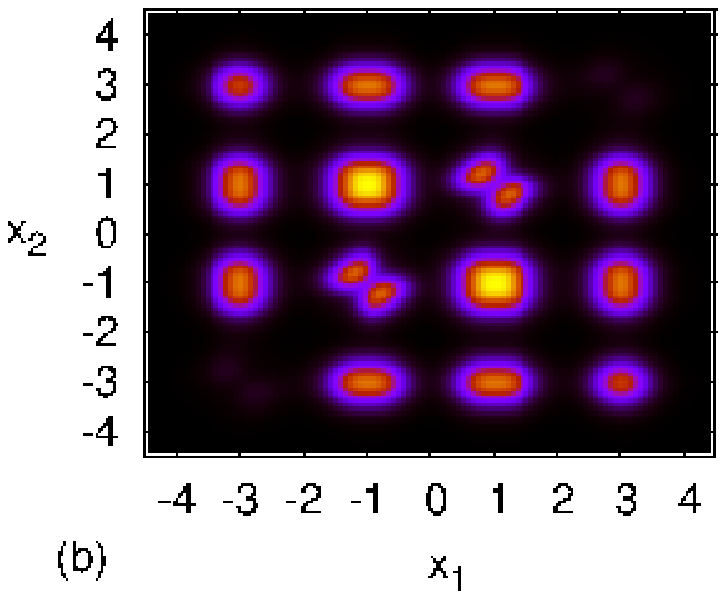} \\
      \includegraphics[width=4.0 cm,height=4.0 cm]{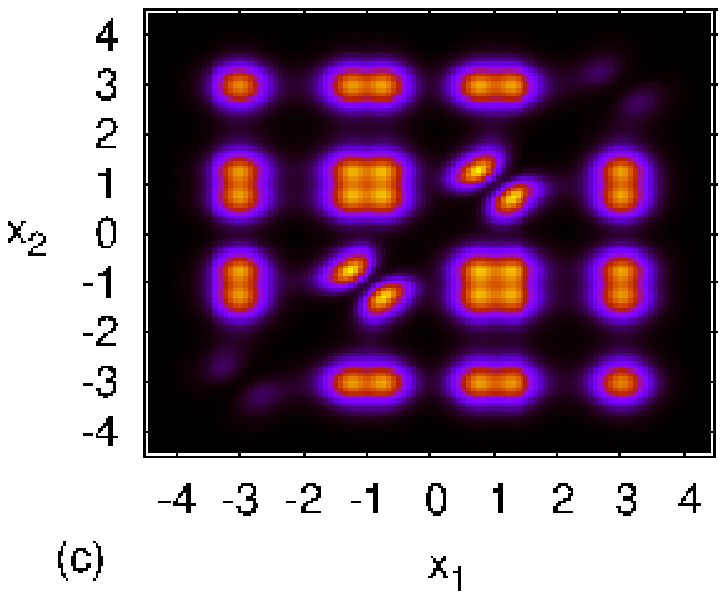} &
      \includegraphics[width=4.0 cm,height=4.0 cm]{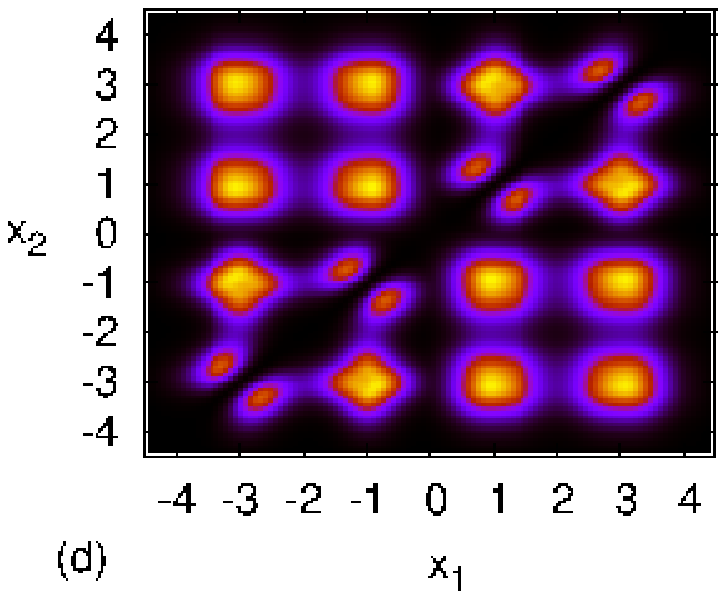} \\
    \end{tabular}
\caption{(Colour online) Two body-density $\rho_2(x_1,x_2)$ for 6 particles and 4 wells for (a) $g=0.1$, (b) $g=10.0$, (c) $g=30.0$, (d) $g=100.0$.}
  \label{fig15}
  \end{center}
\end{figure}

Let us proceed by analysing the case of 6 particles in 4 wells from a two-body perspective. 
For weak interactions, the off-diagonal peaks of $\rho_2(x_1,x_2)$ become more pronounced than  the diagonal contribution (Fig.~\ref{fig15} $g=0.1$). 
For stronger interactions ($g=10.0$) we observe reconcentration in the center and the correlation hole is starting to be formed. 
The off-diagonal humps start to broaden in the middle and acquire fragmented patterns while the correlation hole becomes more evident ($g=30.0$). 
In the fermionization limit ($g=100.0$) the distribution on the diagonal becomes equal for all sites, and the off-diagonal contributions lose their fragmented pattern. 
As a last comment for the comparison of the crossover seen in the one- and two-body density,
 we observe that the formation of correlation hole in the diagonal of $\rho_2(x_1,x_2)$ (Fig.~\ref{fig15} $g=10.0$) corresponds to a widening of the corresponding peaks in $\rho(x)$ 
 (Fig.~\ref{fig14}(a) $g=10.0$), while the appearance of wiggles (Fig.~\ref{fig14}(a) $g=30.0$) is connected with the fragmented patterns in the corresponding off-diagonal humps of $\rho_2(x_1,x_2)$
 (Fig.~\ref{fig15} $g=30.0$).

\begin{figure}

	\includegraphics[width=2.7 cm,height=2.7 cm]{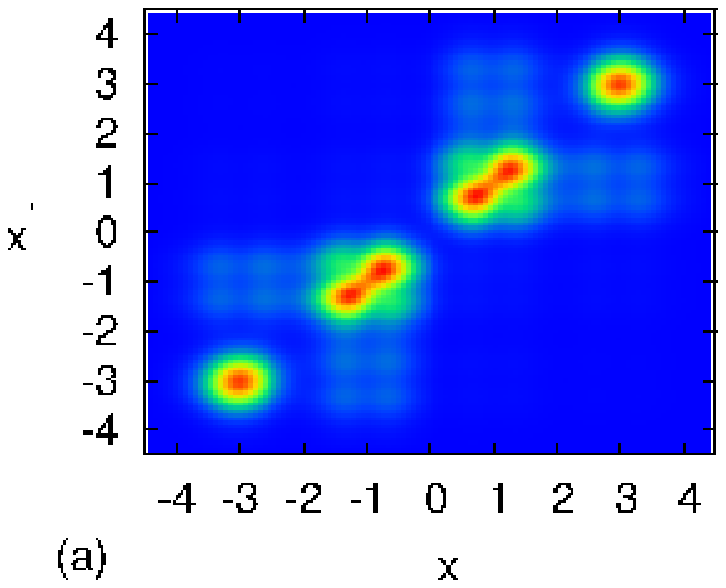} 
	\includegraphics[width=2.7 cm,height=2.7 cm]{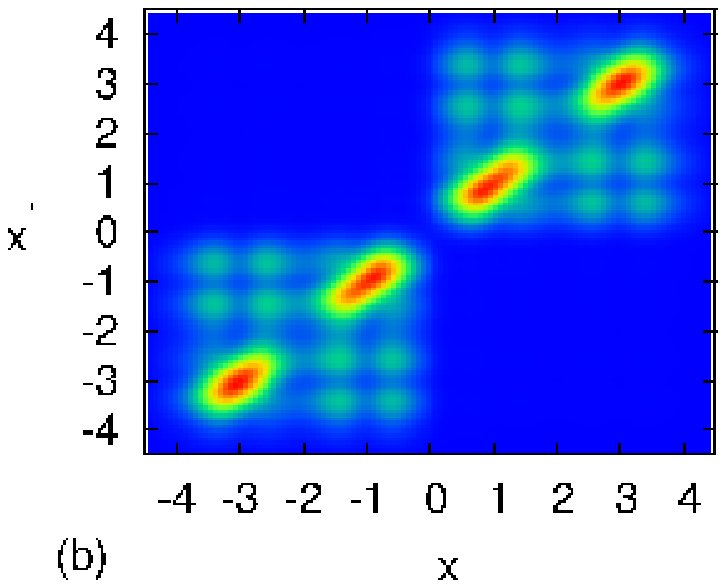} 
	\includegraphics[width=2.7 cm,height=2.7 cm]{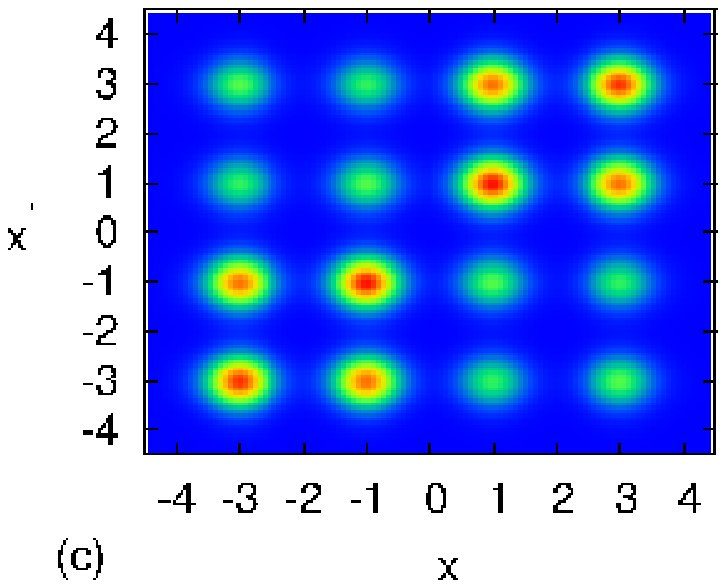} 
	\includegraphics[width=8.6 cm,height=4.3 cm]{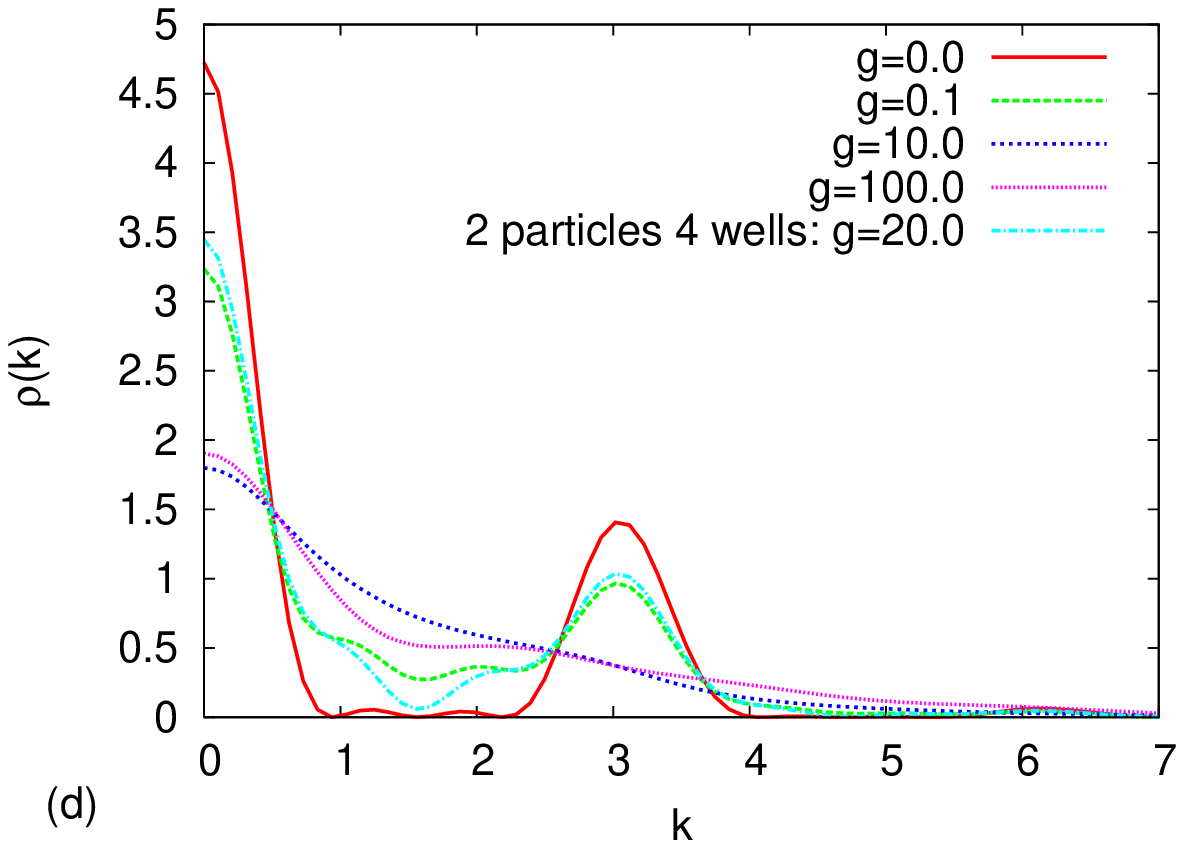}  	

\caption{(Colour online) One-body density matrix $\rho(x,x')$ for 6 particles in 4 wells (a) $g=30.0$, (b) $g=100$. and (c) 2 particles in 4 wells fermionization limit $g=20.0$. (d) Momentum distribution for 6 particles in 4 wells. Shown are 6 different values of $g$: non interacting ($g=0.0$), weak interacting ($g=0.005,0.1, 2.0$), strong interactions ($g=10.0, 30.0$) and fermionization limit ($g=100.0$) which is compared with the fermionization limit of 2 particles in 4 wells.}
  \label{fig16}

\end{figure}

An interesting effect occurs in this case for the one-body density matrix  $\rho_1(x,x')$. 
The remaining coherence in the strongly interacting limit is only concentrated in the left and the right part of the space (diagonal squares in Fig.~\ref{fig16} (a),(b)). 
This indicates that the two extra particles, tend to localise with respect to each other since the off-diagonal squares of this plot are completely depleted;
 the left is completely uncorrelated with the right part of the space. 
Apparent is the formation of wiggles in the center of the diagonal for $g=30.0$ (Fig.~\ref{fig16}(a))
 along with the formation of fragmented patterns in the remaining off-diagonal contribution. 
For the fermionization limit (Fig.~\ref{fig16}(b) $g=100.0$) these wiggles are smeared out to an equally broadened profile,
 while there is a very strongly fragmented pattern in the populated off-diagonal spots. 
It is worth comparing this to the fermionization limit with the corresponding $\nu^<=2/4$ case, 2 particles in 4 wells (Fig.~\ref{fig16} (c)). 
The coherence between left and right part of the potential is present in the latter case (off-diagonal squares retain contribution),
 which means that one particle mainly localised in the left part does penetrate into the right part. 
Contrarily, the background of localised particles in the case of 6 particles in 4 wells,
 prevents each of the excited extra particles from intruding into the side where the other one sits. 

The localisation of the extra particles in the two middle wells for strong interactions leads to an almost smoothened (MI) profile of the momentum distribution
 (Fig.~\ref{fig16}(d) $g= 10.0$). 
As we go to the fermionization limit ($g=100.0$), the central peak becomes a little higher, and the profile is somewhat distorted again,
 as the coherence slightly increases because of the delocalisation of the extra particles. 
Of course the corresponding $\nu^<=2/4$ case (Fig.~\ref{fig16}(d) 2 particles in 4 wells) is much less smoothened,
 since the delocalisation is not hindered by any localised background.

\section{Conclusions and Outlook}

We have performed a numerically exact investigation of few-boson systems in finite one-dimensional lattices for varying strength of the repulsive interactions. 
In the unit filling case, we have shown the evolution from a non-interacting state with a maximum of the density in the center to an equal site distribution
 and a simultaneous decrease of the fluctuations;
 these effects are in accordance with the predictions of the Bose Hubbard model, following the Superfluid - Mott insulator transition from a delocalised to a localised state. 
Beyond that, for higher commensurate filling, on-site interaction effects like broadening and wiggles in the one-body density
 as well as correlation hole and fragmented patterns in the two body density occur especially in the strongly repulsive limit approaching fermionization. 
The coherence loss due to the interaction is reflected in the reduced off-diagonal contribution in the one-body density matrix
 and the smoothening of the peaked pattern in the momentum distribution. 
The high-momentum tails show a difference between the Mott-insulator within the Bose-Hubbard model with unperturbed on-site functions
 and the Tonks limit with two fermionized particles per site. 
Interestingly, there is a slight increase of long-range correlations on the onset of interactions with the particles remaining mostly in one orbital which broadens. 
The fragmentation into different natural orbitals for strong interactions reflects the band structure of the lattice. 
The effect of a deeper lattice on enhancing and accelerating these processes was pointed out. 
For incommensurate filling we have shown that the density distributes inhomogeneously depending on the number of particles and wells,
 and partially delocalised incoherent states are formed. In particular for filling factor greater than one,
 the degree of localisation of these extra particles depends on the interaction strength, the presence of the background particles as well as the specific conditions of the setup. 
For finite systems, it can vary locally, leading to spatial variations for the observables,
 in analogy with the insulating and superfluid domains which appear in the case of an additional harmonic confinement. 
Our study suggests many promising routes for further investigations, like spatially inhomogeneous lattices,
 particularly the case of an external harmonic trap and disorder with arbitrary energy offsets between the wells. 

\acknowledgments

The authors would like to thank Hans-Dieter Meyer, Elmar Haller and Hans-Christoph N\"agerl for helpful discussions. I.B. also acknowledges Lincoln Carr and S.Z acknowledges T. Schuster. 

\appendix
\section{Computational Method}

The goal is to investigate the ground state properties of finite bosonic systems introduced in Sec. II for the complete range of interaction strengths in a numerically exact way. 
Our approach relies on the Multi-Configurational Time-Dependent Hartree (MCTDH) method \cite{mctdhbook,meyer90,beck00},
 primarily a wave-packet dynamics tool known for its outstanding efficiency in high-dimensional applications. 
The underlying idea of MCTDH is to solve the time-dependent Schr\"odinger equation

\begin{equation}
\left\{ \begin{array}{c}
i\dot{\Psi}=H\Psi\\
\Psi(Q,0)=\Psi_{0}(Q)\end{array}\right.\label{eq:TDSE}
\end{equation}
as an initial-value problem by expansion in terms of direct (or Hartree)products $\Phi_{J}$:

\begin{eqnarray}
\nonumber
\Psi(Q,t)&=&\sum_{J}A_{J}(t)\Phi_{J}(Q,t)
\nonumber\\
&\equiv&\sum_{j_{1}=1}^{n_{1}}\ldots\sum_{j_{f}=1}^{n_{f}}A_{j_{1}\ldots j_{f}}(t)\prod_{\kappa=1}^{f}\varphi_{j_{\kappa}}^{(\kappa)}(x_{\kappa},t),\label{eq:mctdh-ansatz}
\end{eqnarray}
using a convenient multi-index notation for the configurations, $J=(j_{1}\dots j_{f})$, where $f=N$ denotes the number of degrees of freedom and $Q\equiv(x_{1},\dots,x_{f})^{T}$. 
The \emph{single-particle functions} $\varphi_{j_{\kappa}}^{(\kappa)}$ are in turn represented in a fixed, primitive basis implemented on a grid. 
For indistinguishable particles as in our case, the single-particle
functions for each degree of freedom $\kappa=1,\dots,N$ are of course identical in both type and number ($\varphi_{j_{\kappa}}$, with $j_{\kappa}\le n$).

In the above expansion, both the coefficients $A_{J}$ and the Hartree products $\Phi_{J}$ are time-dependent. 
Using the Dirac-Frenkel variational principle, one can derive equations of motion for both $A_{J},\Phi_{J}$. 
This conceptual complication offers an enormous advantage: the basis $\{\Phi_{J}(\cdot,t)\}$ is variationally optimal at each time $t$, allowing us to keep it fairly small. 
The permutation symmetry can be enforced by symmetrising the coefficients $A_{J}$, but the ground state is automatically bosonic. 

The Heidelberg MCTDH package  \cite{mctdh:package}, incorporates the so-called \emph{relaxation method}
 which provides a way to obtain the lowest \emph{eigenstates} of the system by propagating some wave function $\Psi_{0}$ by the non-unitary $e^{-H\tau}$ (\emph{propagation in imaginary time}.) 
As $\tau\to\infty$, this automatically damps out any contribution but that stemming from the true ground state $|\mathbf{0}\rangle$,
\[ e^{-H\tau}\Psi_{0}=\sum_{J}e^{-E_{J}\tau}|J\rangle\langle J|\Psi_{0}\rangle.\]  
In practice, one relies on a more sophisticated scheme termed \emph{improved relaxation} \cite{meyer03}. 
Here $\langle\Psi|H-E|\Psi\rangle$ is minimised with respect to both the coefficients $A_{J}$ and the configurations $\Phi_{J}$. 
The equations of motion are solved iteratively, first for $A_{J}(t)$
 (by diagonalisation of $(\langle\Phi_{J}|H|\Phi_{K}\rangle)$ with fixed $\Phi_{J}$) and then propagating $\Phi_{J}$ in imaginary time over a short period. 
The cycle will then be repeated.

As it stands, the effort of this method scales exponentially with the number of degrees of freedom, $n^{N}$. 
Just as an illustration, using $15$ orbitals and $N=5$ requires $7.6\cdot10^{5}$ configurations $J$. 
This restricts our analysis in the current setup to about $N=O(10)$, depending on how decisive correlation effects are. 
If these are indeed essential, then it turns out that at least $n=N$ orbitals are needed for qualitative convergence alone, while the true behaviour may necessitate about $15$ for $N=5$. 
By contrast, the dependence on the primitive basis, and thus on the grid points, is not as severe. 
In our case, the grid spacing should of course be small enough to sample the interaction potential,
 and we consider a basis set of sinusoidal functions which guarantee the hard-wall boundary condition (zero value of the wave function on the first and the last grid point). 
Of interest from a methodological point of view (but also from a conceptual one) is that a shallower lattice enhances the convergence of this computational method. 
For a deep lattice the localisation of the particles in individual wells is pronounced and thus a higher number of delocalised single-particle functions are necessary to express them.

\end{document}